\documentclass[11pt]{article}
\usepackage{graphicx}
\usepackage{times}
\usepackage{amsmath,amsthm,amssymb,setspace,enumitem,epsfig,titlesec,verbatim,color,array,eurosym,multirow}
\usepackage[sort&compress,comma,round,numbers]{natbib}
\usepackage[small,bf]{caption}
\usepackage[margin=2.5cm, includefoot, footskip=30pt]{geometry}
\usepackage{standalone}
\usepackage{tikz}
\usepackage{hyperref}
\usepackage{tabularx}
\usepackage{booktabs}
\usepackage{blkarray}
\usepackage{lineno}
\smallskip

\def\Methods{\textbf{Methods}}

\titleformat{\section}{\sffamily \fontsize{12}{14}\bfseries}{\thesection}{1em}{}
\titleformat{\subsection}{\sffamily \fontsize{11.5}{11.5}\bfseries}{\thesubsection}{1em}{}

\newcounter{extendedfigure}

\title{\bf  \sffamily \Large Simple evolution drives direct reciprocity to maximum payoff in social dilemmas\\}
\date{}
\author{
  \parbox[c]{14cm}{
    \centering
    \onehalfspacing
    Martin A. Nowak\\[1.5ex]
    \raggedright
        Department of Mathematics, Department of Organismic and Evolutionary Biology, Harvard University, Cambridge, MA 02138, USA\\
     }
}

\usepackage{amsmath,amsthm,amssymb,setspace,enumitem,epsfig,titlesec,verbatim,color,array,eurosym,multirow}
\usepackage{subcaption}
\usepackage[all]{hypcap}

\newcommand{\SWAP}{\textrm{SWAP}^{*}}
\newcommand{\HOLD}{\textrm{HOLD}^{*}}
\newcommand{\Ran}{\textrm{RAN}}
\newcommand{\eps}{\epsilon}
\newcommand{\Ex}{\mathbb{E}}

\begin{document}

\maketitle

~\\[0.4cm]
\noindent
{\bf
Direct reciprocity is a mechanism for evolution of cooperation based on repeated interactions between the same individuals. Direct reciprocity can help natural selection to favor cooperators over defectors, but whether or not cooperation prevails depends on the details of the evolutionary dynamics. We describe simple processes of evolution that have the astonishing ability of driving direct reciprocity to maximum payoff in all social dilemmas which we study. The basic process is based on mutation and pairwise comparison. Mutation samples strategies near the boundary of the strategy space. Pairwise comparison includes a parameter for intensity of selection. For large population sizes, intermediate to high mutation rates and intermediate to strong intensities of selection, we find that the process leads to communities of strategies that reach maximum payoff in Prisoner's Dilemma, Snowdrift, Stag Hunt and Harmony games. Maximum payoff in all four games is consistently achieved if players have access to memory-2 strategies. Memory-1 strategies have the capacity to resolve all four social dilemmas, but they are usually defeated by a ``Hold-trap'' in Snowdrift games.}

\newpage

\noindent

\noindent Cooperation and defection are fundamental aspects of the biological world and of human society~\citep{rapoport1965prisoner,trivers:QRB:1971,Axelrod:Science:1981,nowak:Science:2006,Sigmund:2010, Rand:2013,Spadaro:PPS:2022}. 
Many of our attempts succeed or fail because individuals cooperate or defect.
For humans, cooperation is decisive on the small scale of building successful teams as well as on the large scale of establishing sensible societies, productive relationships between countries, global cooperation and cooperation with future generations~\citep{olson:book:1971,milinski:Nature:2002,hauser:Nature:2014,Frank:SciAdv:2018,Gokhale:HSSC:2022}.
Cooperation means paying a cost for someone else to have a benefit. The inherent problem of cooperation is that it is opposed by selfishness. Why should you help an individual who is a potential competitor? Natural selection frowns on cooperation unless a mechanism is in place that favors cooperation over defection~\citep{nowak:Science:2006}.
A mechanism is an interaction structure within a population that improves the chances of cooperators to prevail. For humans, two important mechanisms are direct reciprocity~\cite{trivers:QRB:1971,Axelrod:Science:1981,Nowak:Nature:1992,Nowak:Nature:1993,Hauert:ProcB:1997,killingback:AmNat:2002,szolnoki:pre:2009b,van-segbroeck:PRL:2012,akin2012, Press:PNAS:2012,Stewart:PNAS:2013,McAvoy:PNAS:2016,Hilbe:NHB:2018,Murase:ScientificReports:2020, Glynatsi:HSSC:2021,chen:PNASnexus:2023,tkadlec2023m,Glynatsi:PNAS:2024,Rossetti:Ethology:2024}, which is based on repetition, 
and indirect reciprocity~\cite{nowak:Nature:1998, Leimar:ProcB:2001,  Ohtsuki:JTB:2004, Nowak:Nature:2005,Sigmund:JTB:2012,Santos:Nature:2018, Okada:ScientificReports:2018, radzvilavicius:elife:2019,Gross:NComms:2019,Fujimoto:PNAS:2023,Cavaliere:SciRep:2024,Murase:PNAS:2024a, Michel-Mata:Nature:2024,Wei:JRSI:2025}, which is based on reputation. \\

\noindent In the standard formulation of direct reciprocity, there are repeated interactions between the same two individuals. In any one round of the game, I am tempted to defect to earn a slightly higher payoff. But if I defect today, my co-player may reciprocate with defection tomorrow. If I cooperate today, my co-player may cooperate tomorrow. The light of the future is cast over the present moment. If the game is repeated long enough, then  - according to the Folk theorem - any equilibrium payoff can be reached as long as both players get more than the payoff for mutual defection. \\

\noindent But whether or not cooperation is achieved depends on the details of the evolutionary process and the parameters of the game. Cooperation is never fully stable~\cite{boyd:Nature:1987}. Often we witness cycles of cooperation and defection~\cite{nowak1989osc,imhof:PNAS:2005,Garcia:Frontiers:2018}. Again and again cooperation falters, followed by long periods of defection, which require a painful re-building of cooperation. In many studies of the donation game, which is a simplified Prisoner's Dilemma, low cost-to-benefit ratios are required for evolutionary dynamics to generate high levels of cooperation~\cite{imhof:PRSB:2010,Baek:SciRep:2016}. The exception is~\cite{tkadlec2023m} who show that mutation enhances cooperation in the donation game.\\

\noindent The basic idea of evolutionary game theory is that individuals adopt strategies, that they receive payoff by interacting with others and that payoff contributes to reproductive success~\cite{JMS82,weibull1997,hofsig1998,nowak2006book,broom2022book}. Here we describe a simple process of evolutionary game dynamics that has the ability to find maximum payoff in all social dilemmas which we study. We focus on two-person, two-action games such as the Prisoner's Dilemma (PD), the Snowdrift (SD) game, the Stag Hunt (SH) game~\cite{skyrms2004stag} and the Harmony (HY) game. We explore the repeated version of those games. In any one round, we have the following relationship between cooperation, $C$, and defection, $D$: 
In the PD game, cooperation is dominated by defection.  
In the SD game, I maximize my payoff by doing the opposite as my co-player. 
In the SH game, I maximize my payoff by doing the same as my co-player. 
In the HY game, cooperation dominates defection.
The PD is the most stringent social dilemma, while the SD and SH games and some HY games are relaxed social dilemmas~\cite{nowak:JTB:2012}. Once we turn to the repeated versions of those games an interesting subtlety arises. In some repeated PD, SD and HY games, alternation between cooperation and defection is awarded a higher payoff than full cooperation: two players who alternate between $CD$ and $DC$ rounds achieve a higher payoff than two players who only play $CC$ rounds. Curiously, the evolutionary processes, which we are about to describe, can also resolve those games to maximum payoff. The evolutionary dynamics either find full cooperation or perfect alternation whatever is payoff maximizing.\\ 

\noindent
\textbf{Strategies.} In each round of the repeated game there are four possibilities: $CC, CD, DC, DD$. The players use strategies that depend on the outcome of previous rounds. A memory-n strategy takes into account the outcome of the last $n$ rounds.  A memory-$n$ strategy is given by a vector of $4^n$ probabilities. Each component of the vector specifies for a particular history of the last $n$ rounds the probability to cooperate in the next round. The game between two players that use memory-$n$ strategies leads to a Markov chain on $4^n$ states. If all probabilities, which specify the strategies of both players, are taken from the open interval $(0,1)$ then the transition matrix of the Markov chain is mixing and the stationary distribution of the repeated game is well-defined. The per-round payoff is  calculated from the stationary distribution (see~\Methods{}).\\

\noindent Memory-1 (M1) strategies~\cite{Nowak:Nature:1993} are given by 4 components, $(p_1,p_2,p_3,p_4)$, which denote the probabilities to cooperate given the outcome of the previous round was $(CC, CD, DC, DD)$. M1 strategies include:  always defect, ALLD $(0,0,0,0)$, always cooperate, ALLC $(1,1,1,1)$, tit-for-tat, TFT $(1,0,1,0)$, generous-tit-for-tat, GTFT $(1,x,1,x)$ with $x>0$, or win-stay, lose-shift, WSLS $(1,0,0,1)$. The space of M1 strategies is the hypercube $[0,1]^4$. Note that ALLC and ALLD belong to the subset of memory-0 strategies, while TFT and GTFT belong to the subset of reactive strategies~\cite{Nowak:Nature:1992}, which can also be designated as memory-1/2.
Memory-2 (M2) strategies are given by 16 components, $(p_1,...,p_{16})$. 
See~\Methods{} for notation and details. \\

\noindent \textbf{Two person social dilemmas.} All symmetric, two-person, two-action social 
dilemmas~\citep{kerr:TREE:2004,nowak:JTB:2012,Pena:DGA:2023} can be represented by the following payoff matrix for actions $C$ and $D$:
\begin{equation} \label{E1}
    \bordermatrix { & C & D \cr
                    C & 1 & u \cr
                    D &1+v & 0}                    
\end{equation}
If both players cooperate, they get payoff 1. 
If both players defect, they get payoff 0. 
If one player cooperates, while the other defects, the cooperator gets $u$ and the defector gets $1+v$.
These games include the PD game ($u\!<\!0$, $v\!>\!0$),
the SD game ($u\!>\!0$, $v\!>\!0$), 
the SH game ($u\!<\!0$, $v\!<\!0$), 
and the HY game ($u\!>\!0$, $v\!<\!0$). 
The four games form the quadrants on the $u,v$-plane.
The origin is the game $(u,v)=(0,0)$, which represents  the contact point of all four dilemmas.\\

\noindent For most of the paper, we study infinitely repeated games. We calculate the average payoff per round.
If $u\!+\!v < 1$ then the maximum payoff is achieved by full cooperation.
If $u\!+\!v > 1$ the maximum payoff is achieved when the two players alternate between $CD$ and $DC$.
For memory-n strategies (with $n \geq 1$), the maximum and minimum per round payoffs that can be achieved 
in the infinitely repeated game are given by $\pi_{\max}= \max \{ 1, (1+v+u)/2\}$ and 
$\pi_{\min}= \min \{ 0, (1+v+u)/2\}$. Throughout the paper, we report the scaled average payoff as 
$\bar\pi_s = (\bar \pi -\pi_{min})/ (\pi_{\max}-\pi_{\min})$, where $\bar \pi$ is the long-term average payoff in a population. We can say that $\bar\pi_s$ is the efficiency of an evolutionary process to resolve the particular game. For the donation game, $\bar\pi_s$ has a particularly simple meaning: it is exactly the average cooperation rate of the population (see SI).\\

\noindent In all SH games, the maximum payoff is awarded to strategies that achieve full cooperation.
The three other quadrants, containing PD, SD  and HY, are split by the line $u\!+\!v = 1$ into two areas. Below this line, maximum payoff is awarded to full cooperation. Above this line, maximum payoff is awarded to alternators. This subtlety is often avoided in the literature of the PD by focusing on those games that are below the line. The donation game, which is given by $u\!+\!v = 0$ with $v>0$, is entirely below the line.\\

\noindent There is a small triangle of SD games, which is below the line $u\!+\!v = 1$. In all other SD games, maximum payoff is awarded to strategies that opt for alternation. There are two possibilities to achieve alternation: SWAP strategies alternate after every round, while HOLD strategies alternate over long periods. A memory-1 SWAP strategy is $(1/2, 0, 1, 1/2)$. A memory-1 HOLD strategy is $(1/2, 1, 0, 1/2)$. Both of those strategies achieve maximum payoff in games with $u+v>1$.\\

\noindent \textbf{Evolutionary dynamics.}  In a population of size $N$,  each individual uses a strategy for playing a repeated game. Individuals receive payoffs from interactions with others. Our basic model is similar to a frequency dependent Wright-Fisher process.  First, there is an interaction phase that leads to payoffs for all individuals.   Each individual chooses $m$ consecutive co-players for pairwise interactions. All payoffs, which an individual receives, are added up to represent the total payoff of that individual. The expected number of interactions for each individual is $2m$, but the actual number of interactions varies stochastically between individuals. For updating we consider the total payoffs of each individual, although some individuals played more games than others. For computational efficiency, we choose $m=1$. Increasing $m$ does not change our results.\\

\noindent After all interactions have occurred there is an updating phase. Each individual has the option to update their strategy. With probability, $\mu$, representing mutation, the individual chooses a random strategy from the strategy space. With probability, $1-\mu$, the individual chooses a random role model for pairwise comparison~\cite{Traulsen:JTB:2007}. The focal individual adopts the role model's strategy with probability $1 / [1+\exp [-\beta(\pi_R-\pi_F)/(2m)]]$. Here $\pi_F$ and $\pi_R$ are the payoffs of focal player and role model that were obtained in the interaction phase. The parameter $\beta$, which is non-negative, denotes intensity of selection. The process is followed over a long period of time, and scaled average payoffs (efficiencies) are reported (as explained above).\\

\noindent \textbf{Mutation.} 
The most relevant M1 or M2  strategies reside near the boundary of the strategy space, which means that some of the components of their strategy vector, $\vec p=(p_1,...,p_n)$, are close to 0 or 1. It is useful to consider mutation that biases $p_i$ toward  0 and 1. But it is best to avoid $p_i$ that are exactly 0 or 1, because this choice would imply a world without behavioral noise - without trembling hands. Moreover, the game between some deterministic strategies does not lead to a unique stationary distribution, but depends on initial moves. We consider two options: (1) When studying the space of continuous M1 or M2 strategies we generate mutants as follows: with probability $\alpha$ we set $p_i=p_{min}$; with probability $\alpha$ we set $p_i=p_{max}$; with probability $1-2\alpha$ the value of $p_i$ is uniformly distributed between $p_{min}$ and $p_{max}$. We set $p_{min}=\epsilon$ and $p_{max}=1-\epsilon$, where $\epsilon$ is the noise floor. For many of our simulations, we use $\alpha=1/3$ and $\epsilon=10^{-4}$. 
(2) We study the space of binary strategies, $p_i \in \{ p_{min}, p_{max} \}$, which is equivalent to setting $\alpha=1/2$.\\ 

\noindent \textbf{Previous work.} 
Tkadlec et al~\cite{tkadlec2023m} found that mutation enhances cooperation in direct reciprocity. They 
also use a frequency dependent Wright-Fisher process with pairwise comparison, but with deterministic payoff evaluation: all individuals interact with all others. Hence, their simulations are limited to small population size ($N=100$ for M1  and $N=200$ for reactive strategies). They use a mutational process without bias toward the boundary. They only study the donation game.  The present paper differs from~\cite{tkadlec2023m} in the following main points: (i) here all four social dilemmas are studied; (ii) payoff accumulation is stochastic: each individual chooses $m$ interaction partners (with $m<<N$ and often $m=1$); this feature results in an algorithm with a running time that is roughly linear in $N$; (iii) the mutational process includes a bias toward the boundary; (iv) full stochastic M2 strategies are considered; (v) much larger populations are explored (up to $N=10^5$). The crucial observation of the present paper only emerges in populations of size $N=10^3$ or more. See SI for more discussion of previous work.

\bigskip

\noindent {\bf Results}

\bigskip

\noindent {\bf The space of M1 and M2 strategies}.
Figure 1a shows results for the space of M1 strategies. For population size $N=100$ and mutation rate $\mu=10^{-4}$, the evolutionary process finds maximum payoff in many SH games, in HY games below the diagonal $u+v<1$, in some PD games above that diagonal, and in some SD games. But many PD games and especially donation games, $u+v=0$,  remain at low payoff. The situation improves for higher mutation rate, $\mu=10^{-2}$, and becomes much better for large population size, $N=1000$, and high mutation rate, $\mu=10^{-2}$. Here maximum payoff is reached in all PD, SH and HY games, while most SD games do not reach maximum payoff. Figure 1b shows results for the space of M2 strategies. For population size $N=100$, the situation is similar as for M1, but we already see that M2 strategies are better at resolving SD games.  For population size $N=1000$ and $\mu=10^{-2}$, we observe that M2 strategies resolve all games (with a few random exceptions) to maximum payoff.\\ 

\noindent Therefore, evolutionary dynamics in large populations and for high mutation rate lead to maximum payoff in three of the four social dilemmas when using M1 strategies and in all four social dilemmas when using M2 strategies. The observation is remarkable, because many previous studies of direct reciprocity have concluded that favorable payoff choices (such as low cost-to-benefit ratios in the donation game) are required for evolution of cooperation. Those studies typically consider small population size and low mutation rates or even the limit of zero mutation rate. Our first experiment suggests that large populations ($N=1000$) and high mutation rates ($N\mu=10$) empowers the mechanism to resolve all social dilemmas, especially if M2 strategies are available. M1 strategies still suffice to resolve all PD, SH and HY games.\\

\noindent Figure 2 shows mean values over ten independent realizations of the evolutionary process for M2 strategies. All 289 games converge to maximum payoff. For games with $u+v<1$, the process finds ``nice'' M2 strategies that have the ability to correct mistakes: the probability $p_1$ of the strategy vector, $\vec p=(p_1,...,p_{16})$, converges to maximum value.
For games with $u+v>1$, the probability $p_7$ converges to minimum value and $p_{10}$ converges to a maximum value. These are M2 SWAP strategies. The one exception is the half-line $u-v=1$, along which the payoffs for $CD$ and $DC$ are identical. Here holding an asymmetric outcome and swapping between the two asymmetric outcomes earn the same payoff. Selection does not distinguish M2 HOLD from M2 SWAP strategies, and the process converges to neither: $p_7$ and $p_{10}$ settle at intermediate values that differ from one realization to the next, while the payoff remains maximal.\\

\noindent {\bf M1 strategies struggle to resolve SD games.}  The evolutionary simulations of Figure 1 start with random strategies. Let us try other initial conditions. In Figure 3a, the simulations start with random M1 HOLD strategies given by $(x,1-\epsilon,\epsilon,y)$. In Figure 3b, the simulations start with random M1 SWAP strategies given by $(x,\epsilon,1-\epsilon,y)$. Starting with M1 HOLD  strategies leads to the same outcome as starting with M1 random strategies in all games. But starting with M1 SWAP strategies resolves all games. The evolutionary process enables M1 SWAP strategies to hold on to efficient outcomes, but this state is not reached when starting from other initial conditions such as random M1 or M1 HOLD strategies. Irrespective of initial conditions, the evolutionary process with M1 strategies resolves SD games with $u+v<1$.\\

\noindent Let us focus on SD games with $u+v>1$.
Denote by SWAP* the strategy $(1/2,\epsilon,1-\epsilon,1/2)$. SWAP* achieves maximum payoff in a homogeneous population and is a Nash equilibrium for all SD games. Denote by HOLD* the strategy $(1/2,1-\epsilon,\epsilon,1/2)$. HOLD* also achieves maximum payoff in a homogeneous population, but is a Nash equilibrium only for SD games with the knife-edge condition $v=u-1$. For SD games with $v>u-1$, HOLD*  is dominated by other HOLD strategies such as $(\epsilon,1-\epsilon,\epsilon,\epsilon)$ which do not achieve maximum payoff. For $v<u-1$, HOLD* is dominated by other HOLD strategies such $(1-\epsilon,1-\epsilon,\epsilon,1-\epsilon)$ which do not achieve maximum payoff. Therefore, the efficient HOLD* strategy is replaced by inefficient HOLD strategies if $v \neq u-1$. Once the population has settled on those inefficient M1 HOLD strategies the efficient SWAP* strategy cannot invade. We call this the ``Hold-trap''.\\

\noindent  Why is the evolutionary trajectory starting with random M1 strategies biased toward HOLD? The answer may have to do with the expected payoff that SWAP* and HOLD* receive from random M1 strategies. Integrating over the M1 strategy space we find for the expected payoff
\begin{equation} \label{E2}
           E(\hbox{HOLD*},RAN)-   E(\hbox{SWAP*},RAN) = \xi (u+v)        
\end{equation}
The parameter $\xi$ is positive and depends only on $\epsilon$. In the limit $\epsilon\to 0$ we obtain the constant $\xi=0.04011295828\ldots$, which does not depend on any parameters. The difference is therefore positive whenever $u+v>0$, and in particular in every SD game. (See SI for details.)\\

\noindent Numerical simulations suggest that the inefficient  Hold-trap exists for SD games with $u>1$ and $v>0$, with the exception of the half-line $u-v=1$, where HOLD* is itself a Nash equilibrium and achieves maximum payoff. Outside this region all games with $u+v>1$ (which are those PD, SD or HY games that require alternation for realizing maximum payoff) converge to efficient M1 SWAP strategies.\\

\noindent We note that M2 strategies do not have this problem, because there are many versions of M2 HOLD and M2 SWAP strategies that are both stable and achieve maximum payoff. The evolutionary process always finds one of those strategies.\\  

\noindent  {\bf The space of binary M1 and M2 strategies.} 
Figure 4 shows results of evolutionary dynamics (using $N=1000$ and $\mu=10^{-2}$) for the restricted sets of binary M1 and binary M2 strategies. There are $2^4=16$ binary M1 strategies and $2^{16}=65536$ binary M2 strategies. Binary M1 strategies achieve maximum payoff in some SH games and in some HY games, but otherwise they are inefficient. In contrast, binary M2 strategies find maximum payoff in all games.\\

\noindent {\bf Donation games.}  
We now investigate evolutionary dynamics of the donation game in more detail. The goal is to understand why large populations bring about full cooperation, while small populations tend to fail.
For a donation game with high cost-to-benefit ratio (such as $0.95$) the population oscillates between cooperative and defective states. The scaled average fitness of the population,  $\bar\pi_s$, fluctuates between values close to 0 and values close to 1. In order to measure those oscillations, we divide the state space of the process, which is high dimensional and continuous, into three regions: (i) a cooperative state, (ii) an intermediate state, and (iii) a defective state.  For judging transitions between those states we use hysteresis: if $\bar\pi_s$ climbs above 0.9 we enter the cooperative state; if $\bar\pi_s$ falls below 0.7 we leave the cooperative state;  if $\bar\pi_s$ falls below 0.1 we enter the defective state; if $\bar\pi_s$ climbs above 0.3 we leave the defective state; at other times we are in the intermediate state.  This setup avoids counting small fluctuations when evaluating the mean dwell time in each state.\\

\noindent SI Figure 1 shows the average time spent in the cooperative state for three population sizes, $N=100, 500,$ 
$1000$, and three intensities of selection, $\beta =1, 10, 100$, as function of the mutation rate varying from $\mu=10^{-5}$ to $\mu=10^{-1}$. For very low mutation and for very high mutation, there is little cooperation. But for a broad window of intermediate mutation rates, the larger population sizes, $N=500$ and $N=1000$, spend most of their time in the cooperative state for stronger intensities of selection, $\beta=10$ and $\beta=100$.\\ 

\noindent SI Figure 2 shows the average time spent in all three states. For large population size and strong selection, the intermediate state is short lived unless we use extremely high mutation rates which overwhelm selection and trigger an error threshold. For $N=1000$ and $\beta=100$ we have the following situation: for $\mu<10^{-4}$ the population spends some time in the cooperative state, but more time in the defective state and very little time in the intermediate state;  for $0.005 <\mu < 0.05$ the population spends nearly all time in the cooperative state;  for $\mu>0.05$ the population spends nearly all time in the intermediate state. For extremely high selection intensity, $\beta=10^4$, the cooperative window narrows.\\ 

\noindent SI Figure 3 shows the effect of mutation rates on dwell times in the three states.
The defective state is long lived for small mutation rates. We are waiting for mutation to produce TFT-like strategies that invade ALLD by forming invasion clusters~\cite{nowak2004nat}. The supply of those strategies is mutation limited. For all three population sizes, the dwell time in the defective state decreases with mutation rate until an error threshold is reached. The dwell time in the cooperative state behaves differently for small and for large populations. For $N=100$ the dwell time in the cooperative state decreases with mutation rate. For $N=500$ and $N=1000$ the dwell time in the cooperative state stagnates or increases with mutation rate - up to the threshold value when it declines again. The cooperative state is typically lost by drift which causes reciprocators to become too lenient. In larger populations, mutation can lead to a diverse community of strategies that slows down the drift to ALLC. Therefore, increasing the mutation rate in large populations (below the error threshold) increases the mean dwell time in the cooperative state relative to the mean dwell time in the defective state. This effect explains why large population sizes and intermediate (or high) mutation rates are required to find maximum payoff. \\

\noindent We conclude that increasing heterogeneity (below the error threshold) makes large cooperative populations robust but destabilizes small cooperative populations.  Increasing heterogeneity (below the error threshold) destabilizes defective populations.  For both $N=100$ and $N=1000$ there is an intermediate optimum mutation rate, but the peak in average payoff is much higher and much wider for large populations.\\ 

\noindent SI Figure 4 studies the donation game for two cost-to-benefit ratios, $c=0.95$ and $c=0.99$, as a function of selection intensity, $\beta$, and population size, $N$. We find that intermediate values of $\beta$, roughly between $\beta=10$ and $\beta=5000$ maximize efficiency. But we observe that larger $N$ is always (marginally) better - up to the maximum value ($N=10^5$) which was simulated.\\  

\noindent Figure 5 shows the performance of the evolutionary process for donation games with extremely high cost values.
For population sizes $N=10^3$ and $N=10^4$ high cooperation is reached for cost-to-benefit ratios up to 0.997. As cost approaches benefit, the donation game becomes zero-sum and cooperation becomes a pure transfer; $b/c=1$ is the exact point at which the surplus available to reciprocity vanishes. The evolutionary process can therefore approach this boundary but never reach it, and the closer it comes the larger the population needs to be.\\

\noindent {\bf Variations.}  
In the SI, we vary key parameters and modify features of evolutionary dynamics or game dynamics. SI Figure 5 shows variation in mutational bias, $\alpha$, for M1 and M2 strategies. For $\alpha=0$, few games are resolved. Maximum bias, $\alpha=0.5$, implies that only binary strategies are used, which works well for M2 but not for M1 strategies. SI Figure 6 shows systematic variation in the number of interactions, $m$, the intensity of selection, $\beta$, the mutation rate, $\mu$, and the noise floor, $\epsilon$.  Varying $m$ between 1 and 100 has no discernible effect.  A broad range of selection intensities lead to high efficiency; only very high intensities reduce efficiency.  There is a broad window of intermediate mutation rates that lead to high efficiency. Finally, the process is robust with respect to variation in the noise floor as long as it is not too high.\\ 

\noindent SI Figure 7 shows the performance of M2 strategies for extended game grids with payoff parameters, $u$ and $v$, ranging from $-40$ to $+40$. SI Figure 8 studies M1 strategies in finitely repeated games, which can also be interpreted as discounting the future. SI Figure 9, shows that our main findings can be replicated by a different evolutionary process that is based on asynchronous updating.\\

\newpage

\noindent {\bf Conclusion}

\bigskip

\noindent We have described simple evolutionary processes that enable direct reciprocity to resolve all four social dilemmas. The key idea is to use large population size, intermediate/high mutation rates, and intermediate/strong selection. Increasing the mutation rate augments efficiency up to an error threshold where mutation overwhelms selection. Increasing intensity of selection helps up to the limit where the process becomes too deterministic. Some stochasticity is needed to escape from long lived, defective states that have low payoff.\\

\noindent When studying the space of M1 or M2 strategies, each strategy is a point in a 4 or 16 dimensional hypercube. Most of the relevant strategies exist near the boundary of the strategy space. Therefore, whenever mutation introduces new strategies it is important that strategies near the boundaries are sampled. We use mutation that biases toward the boundary. Using instead an unbiased, uniform distribution greatly reduces the efficiency of the process. We also study binary instead of continuous strategy spaces. We observe that binary M2 strategies resolves all games, while binary M1 strategies resolve only a few.\\ 

\noindent  For computational efficiency we use discrete generations with synchronous updating for the basic process. There is an interaction phase followed by a reproductive phase. After all payoffs have been accumulated, all individuals update at the same time. But the same evolutionary dynamics can be found in a process with continuous reproduction and asynchronous updating (SI Figure 9).\\ 

\noindent  Computing large populations over long time periods becomes efficient when dropping the standard convention that payoff is calculated based on every individual interacting with every other individual. We use the following procedure: each individual choses $m$ random consecutive partners. For each interaction, the payoffs of both players are added to their payoff sum. On average, an individual participates in $2m$ interactions. Each individual has at least $m$ interactions. Some individuals have more interactions than others. We use the cumulative payoff for each individual. For most of our simulations, we use $m=1$, which is sufficient. Variation in $m$ has little effect.\\

\noindent In summary, simple, natural, stochastic processes of evolutionary dynamics, when using large population size 
resolve all social dilemmas to maximum payoff for broad windows of  intermediate to high mutation rates and intermediate to high intensity of selection. Memory-1 strategies reliably succeed in all Prisoner's Dilemmas, Stag Hunt and Harmony games, but tend to fail in the Snowdrift game. Memory-2 strategies (and even binary memory-2 strategies) succeed in all four dilemmas. \\[0.5cm]

{
{\setlength{\bibsep}{0\baselineskip}

}}

~\\

\noindent
{\bf Acknowledgments.}  I worked with various AI tools including Claude Opus 5.0, GPT-5.2, Gemini Pro 3.1.\\

\noindent
{\bf Competing interests.}
The author declares no competing interests.\\

\noindent
{\bf Author contributions.}
M.A.N. conceived the study, performed the analysis, made the figures and wrote the text.\\

\noindent
{\bf Code availability.}
All code was written in Fortran and Python.
It is available at: \url{https://github.com/martin-mn/ResolvingSocialDilemmas}\\

\section*{Methods}

\noindent
\textbf{Payoffs.}
We consider two-action repeated games. In each round, two individuals choose whether to cooperate or defect. 
The payoff matrix is
\begin{align}
    \bordermatrix{%
        & C & D \cr
        C &\ R &\ S\  \cr
        D &\ T &\ P\ \cr
    } 
\end{align}
Here \(R\) is the reward for mutual cooperation, \(S\) the sucker's payoff, \(T\) the temptation to defect, and \(P\) the punishment payoff~\cite{Axelrod:Science:1981}. 
We use two parameterizations: (1) In the normalized games, we set \(R = 1\),~\(S = u\),~\(T = 1 \!+\! v\),  \(P = 0\). (2) In the donation game, we set \(R = b \!-\! c\), \(S = -c\), \(T = b\), \(P = 0\).  The parameters \(b \!>\! c \!>\! 0\) represent benefit and cost of cooperation.\\

\noindent {\bf Memory-1 strategies}\cite{Nowak:Nature:1993} are given by a vector with 4 components: \(\mathbf{p}\!=\! (p_{CC}, p_{CD},p_{DC}, p_{DD})\). The entry \(p_{ij}\) refers to the individual's cooperation probability given that the individual's own previous action was \(i\) and the co-player's previous action was \(j\), with $i,j\!\in\!\{C,D\}$.
The repeated game between two players defines a Markov chain~\cite{Nowak:Nature:1993}. 
Let \(\mathbf{p} =(p_{\text{CC}}, p_{\text{CD}}, p_{\text{DC}}, p_{\text{DD}})\) and \(\mathbf{q} = (q_{\text{CC}}, q_{\text{CD}}, q_{\text{DC}}, q_{\text{DD}})\) denote the strategies of players 1 and 2, respectively. The Markov chain has four states, \(\{CC, CD,DC, DD\}\), referring to the outcomes of each round. The transition matrix is
\begin{equation} \label{eq:transition_matrix}
    M = \begin{bmatrix}
    p_{CC} q_{CC} & p_{CC} (1 - q_{CC}) & (1 - p_{CC}) q_{CC} & (1 - p_{CC})(1 - q_{CC}) \\
    p_{CD} q_{DC} & p_{CD} (1 - q_{DC}) & (1 - p_{CD}) q_{DC} & (1 - p_{CD})(1 - q_{DC}) \\
    p_{DC} q_{CD} & p_{DC} (1 - q_{CD}) & (1 - p_{DC}) q_{CD} & (1 - p_{DC})(1 - q_{CD}) \\
    p_{DD} q_{DD} & p_{DD} (1 - q_{DD}) & (1 - p_{DD}) q_{DD} & (1 - p_{DD})(1 - q_{DD})
    \end{bmatrix}
\end{equation}
If all $p$ values are in $(0,1)$ the Markov chain is ergodic and there exists a unique stationary distribution, \(\mathbf{v} = (v_{CC}, v_{CD}, v_{DC}, v_{DD})\), representing the long-run frequencies of the four states. 
This distribution is uniquely determined by the equation \(\mathbf{v} M =\mathbf{v}\), subject to the normalization $v_{CC}\!+\!v_{CD}\!+\!v_{DC}\!+\!v_{DD}\!=\!1$.\\ 

\noindent A standard assumption~\citep{Nowak:Nature:1993,Press:PNAS:2012} is that the game is infinitely repeated and future rounds are not discounted. In this case, the payoffs of player~1 and player~2 are
\begin{equation}
    \pi(\mathbf{p}, \mathbf{q}) = \mathbf{v} \cdot (R, S, T, P) \quad \text{and} \quad \pi(\mathbf{q}, \mathbf{p}) = \mathbf{v} \cdot (R, T, S, P).
\end{equation}
The cooperation rates are \(v_{CC}+v_{CD}\) for player~1 and \(v_{CC} + v_{DC}\) for player~2.\\

\noindent {\bf Memory-2 strategies} are given by a vector with 16 components: 
 $ \mathbf{p}\!=\! (p_{CCCC}, p_{CDCC},p_{DCCC}, p_{DDCC},$   $p_{CCCD}, p_{CDCD},p_{DCCD}, p_{DDCD} , p_{CCDC}, p_{CDDC},p_{DCDC}, p_{DDDC}, p_{CCDD}, p_{CDDD},p_{DCDD}, p_{DDDD} )$. 

\noindent The entry $p_{ijkl}$ refers to my probability to cooperate given my second-to-last move was $i$, your second-to-last move was $j$, my last move was $k$, and your last move was $l$ with $i,j,k,l$ chosen from the set $\{ C, D\}$.
It is useful to label the states 1 to 16 and use the notation $\mathbf{p}\!=\! (p_1, p_2, ..., p_{16})$. The game between two players using M2 strategies defines a Markov chain on 16 states. If all $p$ values are in $(0,1)$ the Markov chain is ergodic and there exists a unique stationary distribution, \(\mathbf{v} = (v_1, ..., v_{16})\), representing the long-run frequencies of the 16 states. The stationary distribution can be calculated by solving a system of linear equations. The payoffs for both players follow from that stationary distribution.\\ 

\noindent \textbf{Evolutionary dynamics.} On a longer timescale, players revise their strategies based on the payoffs they receive. Pairwise interactions are formed to determine payoff. Each individual chooses at random $m$ individuals for a pairwise interaction. Self interactions are not allowed. But an individual could choose the same partner multiple times, because random sampling of partners is done with replacement. Each individual participates at least in $m$ interactions and has on average $2m$ interactions. The number of interactions per individual varies. The payoffs over all interactions of an individual are summed up. That sum is the payoff of that individual. After all interactions have occurred, individuals have the possibility to update their strategies either by mutation or via  pairwise comparison. With probability $\mu$, the focal individual mutates by adopting a random strategy. With probability $1-\mu$, the focal individual randomly selects another individual as role model. The focal individual adopts the role model's strategy with probability $1/[1 + e^{-\beta (\pi_R - \pi_F)/(2m)}]$. The payoffs for focal player and role model are \(\pi_F\) and \(\pi_R\). The selection strength, $\beta \geq 0$, determines sensitivity to payoff differences.\\ 

\bigskip

\clearpage
\begin{center}
\includegraphics[width=\linewidth,height=0.65\textheight,keepaspectratio]{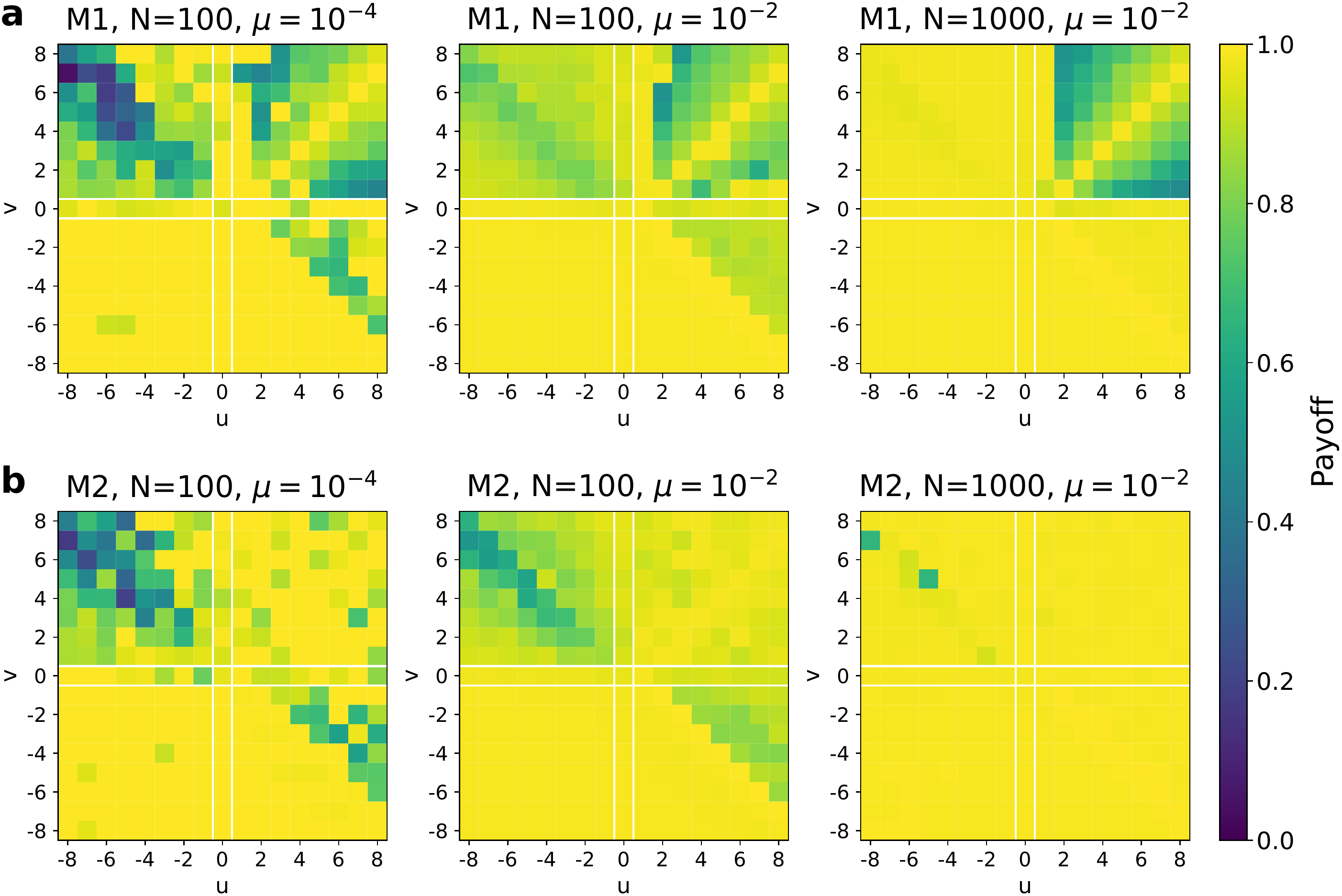}
\end{center}
\vspace{1.5ex}

\noindent {\small {\bf Figure 1: Resolving social dilemmas by direct reciprocity.} The individuals of a populations use strategies to play repeated games. They update their strategies by mutation and pairwise comparison. We study a space of games that includes the Prisoner's dilemma (PD) ($u<0$, $v>0$), the Snowdrift (SD) game  ($u>0$, $v>0$), the Stag Hunt (SH)  game ($u<0$, $v<0$) and the Harmony (HY) game ($u>0$, $v<0$). Population size increases from $N=100$ to $N=1000$ and mutation rate from $\mu=10^{-4}$ to $10^{-2}$. For $N=100$, many PD, SD and HY games remain at low payoff. For $N=1000$ and large mutation rate, memory-2 strategies converge to maximum payoff in almost all games. Memory-1 strategies converge to sub-optimal solutions in some SD games which require alternation ($u+v>1$), but resolve all other games. Parameters: selection intensity, $\beta=10$, mutational bias $\alpha=1/3$, minimum stochasticity $\epsilon= 10^{-4}$, chosen interaction partners $m=1$, time $t=10^6$. Initial condition: random M1 or M2 strategies. See text  for more details.

\bigskip}

\clearpage
\begin{center}
\includegraphics[width=\linewidth,height=0.68\textheight,keepaspectratio]{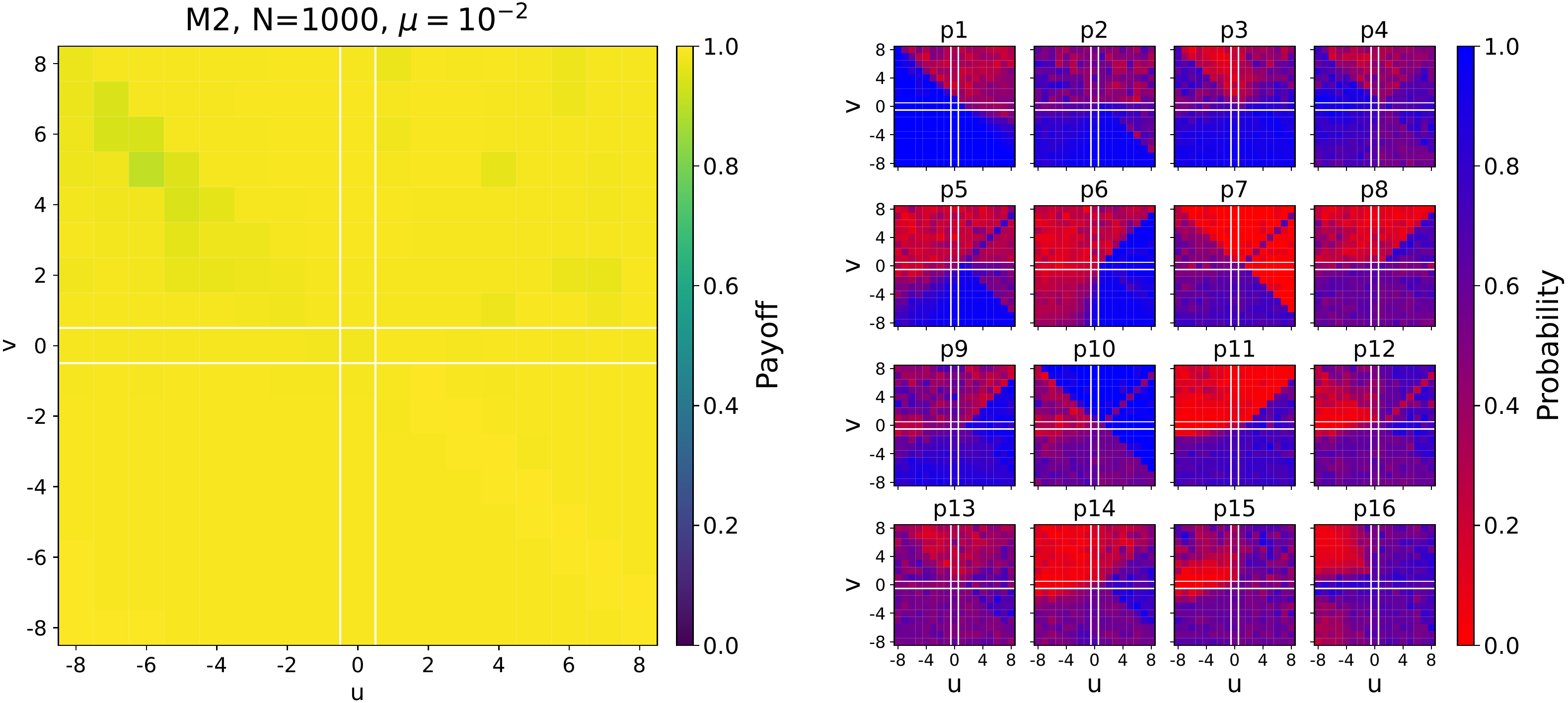}
\end{center}
\vspace{1.5ex}

\noindent {\small {\bf Figure 2: Memory-2 strategies resolve every game to maximum payoff.} Same simulation as in Figure 1, but every value is the mean over 10 runs. The crucial strategic components are: $p_1\approx 1$ for all games with $u+v<1$ and $(p_7,p_{10})\approx (0,1)$ for all games with $u+v>1$ (except when $u-v=1$). Whenever mutual cooperation is optimal, the process finds $nice$ M2 strategies that have the ability to forgive.  Whenever alternation between $CD$ and $DC$ is optimal, the process finds M2 SWAP strategies that are error correcting. On the half line $u-v=1$ (with $u+v>1$) the payoff for $CD$ is the same as the payoff for $DC$. There M2 SWAP and M2 HOLD strategies are payoff-equivalent, the process converges to neither, and $(p_7,p_{10})$ takes intermediate values that vary between runs; the payoff is maximal nonetheless.

\bigskip}

\clearpage
\begin{center}
\includegraphics[width=\linewidth,height=0.67\textheight,keepaspectratio]{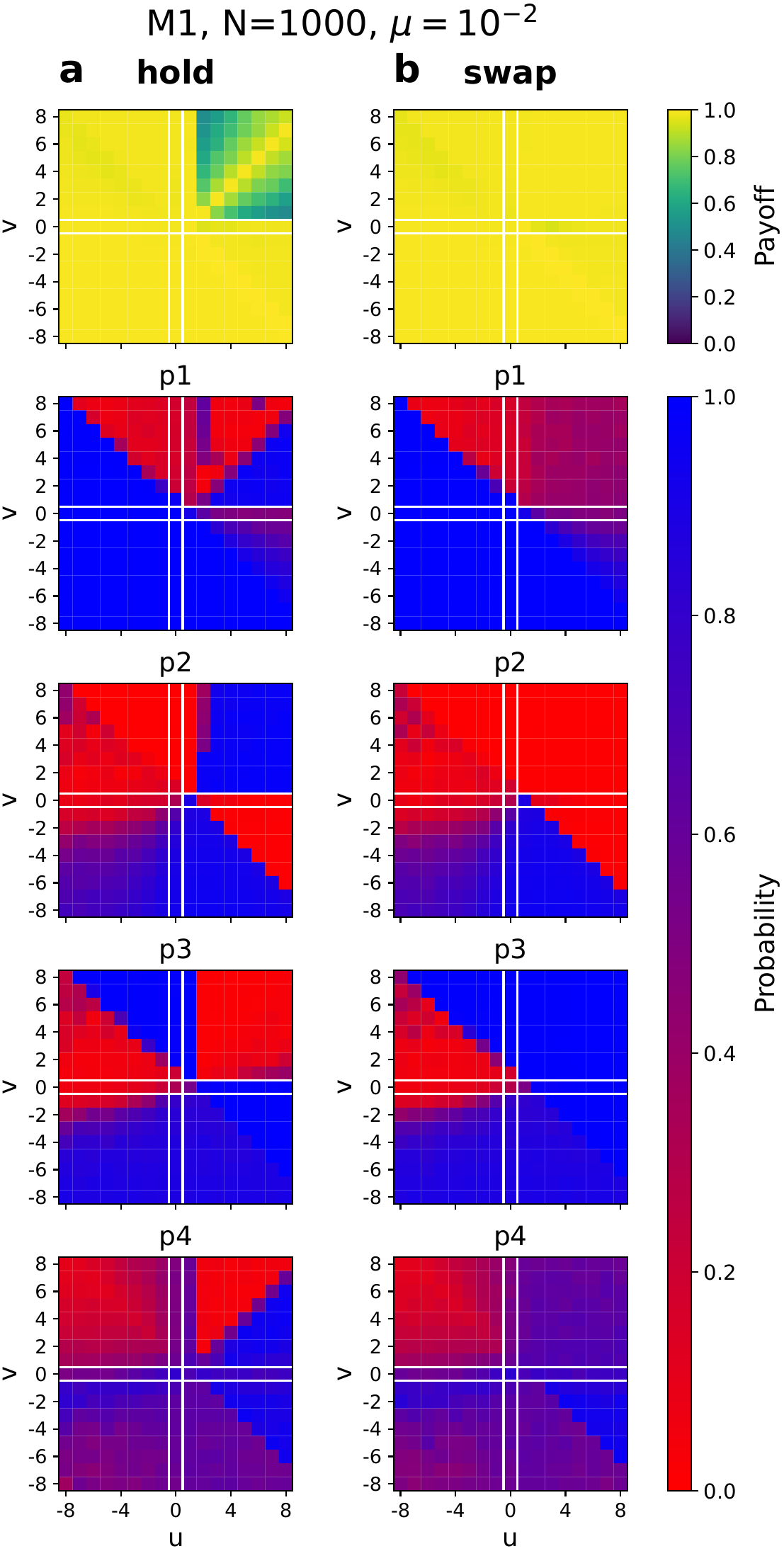}
\end{center}
\vspace{1.5ex}

\noindent {\small {\bf Figure 3: Memory-1 strategies have the capacity to resolve all games.}  On the left, the simulations start with random HOLD strategies $(x, 1,0, y)$. On the right, the simulations start with random SWAP strategies $(x,0,1,y)$. Both initial conditions lead to maximum payoff in all games with $u+v<1$. But in games with $u+v>1$, SWAP strategies hold maximum payoff everywhere, while HOLD strategies converge to sub-optimum solutions for games with $u>1$ and $v>0$ except on the half-line $u-v=1$. For $u-v<1$, the winning Hold strategy is close to $ (0,1,0,0)$. For $u-v>1$, the winning Hold strategy is close to $ (1,1,0,1)$. Neither of those HOLD strategies  achieves maximum payoff. For those games the  Hold strategy  $ (1/2,1,0,1/2)$ would achieve maximum payoff but is not stable. The SWAP strategy $ (1/2,0,1,1/2)$ achieves maximum payoff and is stable, but it is not reached from random initial conditions (see Fig 1) or from random HOLD initial conditions. Same parameters as Fig 1.
\bigskip}

\clearpage
\begin{center}
\includegraphics[width=\linewidth,height=0.68\textheight,keepaspectratio]{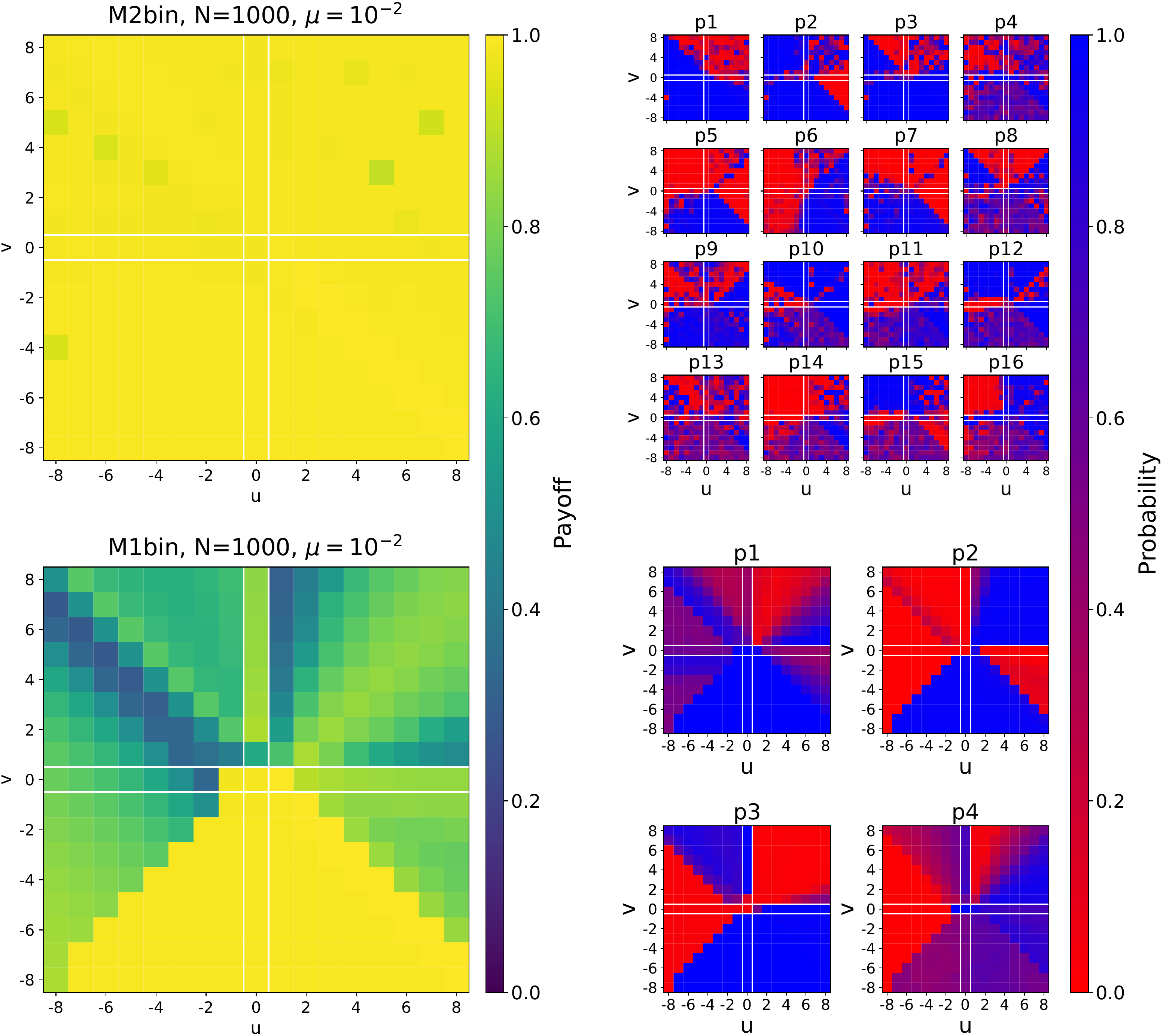}
\end{center}
\vspace{1.5ex}

\noindent {\small {\bf Figure 4: Binary Memory-2 strategies resolve all games, while binary Memory-1 strategies fail.}
There are $2^4=16$ binary M1 strategies and $2^{16}=65536$ binary M2 strategies. We include behavioral  noise by choosing all strategy components, $p_i$, from the set $\{ \epsilon, 1-\epsilon\}$. Binary M2 strategies succeed in resolving all games. In contrast, binary M1 strategies reach maximum payoff only in some SH games with $v \leq u$ and in some HY games with $u+v \leq 1$. In many other games, binary M1 strategies reach intermediate payoff. Same parameters as Fig 1.

\bigskip}

\clearpage
\begin{center}
\includegraphics[width=\linewidth,height=0.68\textheight,keepaspectratio]{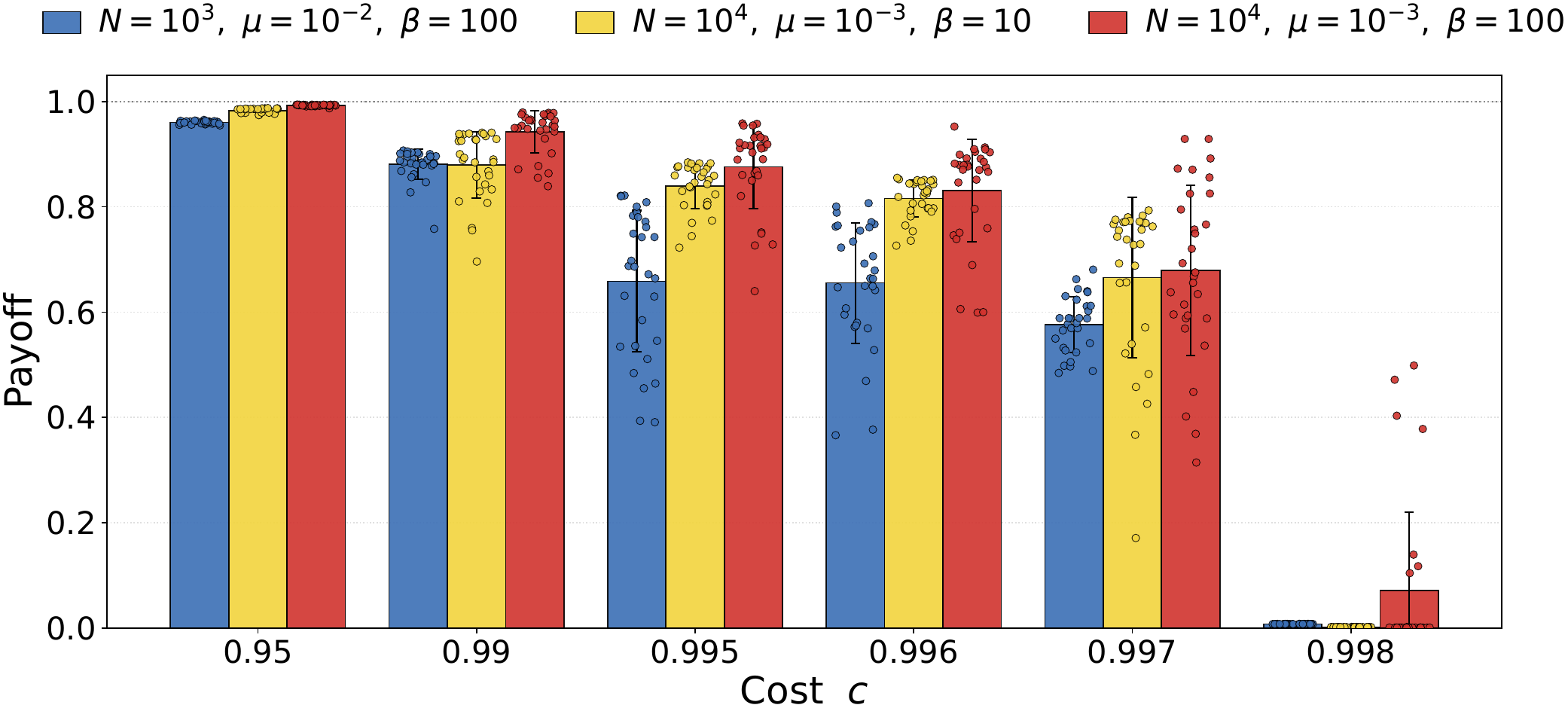}
\end{center}
\vspace{1.5ex}

\noindent {\small {\bf Figure 5: Resolving the donation game for extreme cost values.} We study the donation game for benefit, $b=1$, and cost values, $c$, ranging from 0.95 to 0.998. Three parameter choices are shown: $N=10^3$ with mutation rate $\mu=10^{-2}$ and $\beta=100$; $N=10^4$ with $\mu=10^{-3}$ and $\beta=10$; and $N=10^4$ with $\mu=10^{-3}$ and $\beta=100$. For the same value of $Nu$, larger population size and higher intensity of selection is better for cooperation. The maximum cost value that is compatible with high cooperation rates is $c=0.997$ for the parameters that were used: mutational bias $\alpha=1/3$, minimum stochasticity $\epsilon= 10^{-4}$, chosen interaction partners $m=1$, time $t=10^7$, 30 replicates per run.}

\clearpage
\begin{center}
{\bf  \sffamily \Large Simple evolution drives direct reciprocity to maximum payoff in social dilemmas: Supplementary Information}\\[3ex]
\parbox[c]{14cm}{\centering\onehalfspacing Martin A. Nowak\\[1.5ex]
\raggedright Department of Mathematics, Department of Organismic and Evolutionary Biology, Harvard University, Cambridge, MA 02138, USA\\}
\end{center}
\vspace{2ex}

~\\[1.1cm]

\noindent
The paper investigates simple processes of evolutionary dynamics which are efficient in helping direct reciprocity to resolve social dilemmas. The processes are based on standard assumptions of evolutionary game dynamics. Two main features are: (i) individuals interact randomly to accumulate payoff; (ii) individuals use mutation or pairwise comparison to update their strategies. The crucial observation is: using large population size
and intermediate to strong selection opens a broad window of intermediate to high mutation rates which enable the evolutionary trajectories to converge to maximum payoff in the Prisoner's Dilemma (PD), the Snowdrift game (SD), the Stag Hunt game (SH) and the Harmony game (HY). For donation games, which are special Prisoner's Dilemmas, we find nearly full cooperation even for cost-to-benefit ratios above $0.99$. In principle, donation games for any cost-to-benefit ratio (less than one) can be resolved, but the time for convergence could become very long.\\  

\noindent This SI is in six parts: (i) we give an overview of related literature; (ii) we study evolutionary dynamics of the donation game for high cost values in more detail; (iii) we investigate the robustness of our finding with respect to  parameter variation; (iv) we study finitely repeated games; (v)  we study a similar process with asynchronous updating; (vi) we calculate expected payoff against random M1 strategies.\\

\section{Related literature}\label{SI:section:Literature}

In the following we discuss literature on which this work builds. 

\subsection{Related literature on direct reciprocity}\label{SI:section:direct_reciprocity}

Direct reciprocity is a mechanism for evolution of cooperation~\cite{SI:nowak:Science:2006,SI:Garcia:Frontiers:2018,SI:Glynatsi:HSSC:2021,SI:Rossetti:Ethology:2024}. This mechanism is often studied in the context of the repeated Prisoner's dilemma (PD), although
extensions to other social dilemmas have been considered, including the Snowdrift game~\cite{SI:Kummerli:PRSB:2007}, the Stag Hunt~\cite{SI:Starnini:JSM:2011}, and arbitrary repeated $2\!\times\! 2$ games~\cite{SI:martinez-vaquero:plosone:2012,SI:Stewart:PNAS:2014}.
The PD is particularly interesting, because if played once, classical game theory predicts defection.
In contrast, when interactions are repeated, players can condition their behavior on past outcomes. 
If individuals use strategies such as Tit-for-Tat~\cite[TFT, Ref.][]{SI:Axelrod:Science:1981}, Generous Tit-for-Tat~\cite[GTFT, Refs.][]{SI:molander:jcr:1985,SI:Nowak:Nature:1992}, or Win-Stay, Lose-Shift~\cite[WSLS, Refs.][]{SI:Kraines:TheoryDecision:1989,SI:Nowak:Nature:1993}, mutual cooperation becomes possible.\\

\noindent According to the Folk theorem, repeated games admit a wide range of possible equilibria~\cite{SI:Friedman:RES:1971}. In the repeated PD, these equilibria can give rise to full cooperation, full defection, or alternating cooperation and defection (among many other possibilities) provided each player earns at least the payoff for mutual defection. The question arises: which of these outcomes emerge when populations and strategies evolve over time? Evolutionary game theory provides a natural framework to study this question. Below, we highlight several lines of research that are relevant for our study. \\

\noindent
\textbf{Direct reciprocity with memory-1 strategies.}
Much work on direct reciprocity has focused on memory-1 (M1) strategies, where a player's action depends only on the outcome of the previous round. Within this framework, two complementary approaches have been pursued:
(i) restricting the strategy set to a small number of canonical strategies~\cite{SI:imhof:PNAS:2005,SI:Brandt:JTB:2006,SI:szolnoki:pre:2009b,SI:kurokawa:TPB:2010,SI:Toupo:IJBC:2014,SI:Szolnoki:PRE:2014}, or
(ii) analyzing the full space of M1 strategies~\cite{SI:frean:PRSB:1994,SI:Press:PNAS:2012,SI:Baek:SciRep:2016,SI:Akin:EGADS:2016,SI:chen:PNASnexus:2023,SI:LaPorte:PLoSCB:2023}.\\ 

\noindent While repeated interactions can facilitate evolution of cooperation, cooperation is not guaranteed. 
The evolutionary dynamics depend on the payoff values and on the details of the evolutionary process~\cite{SI:nowak:PNAS:1993,SI:stewart:games:2015,SI:garcia:jet:2016}. 
For large cost-to-benefit ratios $c/b$, cooperation is unlikely to evolve~\cite{SI:Baek:SciRep:2016}.
Here we show how evolutionary dynamics can lead to almost full cooperation even in settings that would normally lead to defection.\\

\noindent
\textbf{All two-person social dilemmas.}
While most work on direct reciprocity considers the repeated PD or the donation game, some studies address the entire space of two-person, two-action games. Martinez-Vaquero et al~\citep{SI:martinez-vaquero:plosone:2012} characterize the outcomes of evolutionary dynamics among M1 strategies over the whole space of symmetric $2\!\times\!2$ games. H\"ubner et al~\citep{SI:Hubner:PNASnexus:2025} generalize the theory of zero-determinant strategies from the donation game to all pairwise social dilemmas, which they define by the condition $\min\{R,T\} > \max\{S,P\}$, where $R$, $S$, $T$ and $P$ denote the payoffs for mutual cooperation, for unilateral cooperation, for unilateral defection and for mutual defection. For each such game they construct equalizer strategies which demonstrate that mutual cooperation can be sustained as an equilibrium, by direct reciprocity, by indirect reciprocity, or by an arbitrary mixture of the two. Two differences to the present work are worth noting. First, their result is a proof of principle: it establishes that the required strategies exist, but it leaves open whether an evolving population would find them. Here we study exactly this question. Second, they identify the socially optimal outcome with mutual cooperation. Their class of games, however, includes games with $T+S>2R$, in which two players who alternate between the two mixed outcomes obtain $(T+S)/2 > R$. In our notation these are the games with $u+v>1$, and they comprise most SD games. In those games mutual cooperation is not payoff maximizing. Our evolutionary process reaches the payoff maximizing outcome in both regimes: it converges to full cooperation if $u+v<1$ and to alternation if $u+v>1$.\\

\noindent
\textbf{Complete strategy spaces.}
When interpreting evolutionary simulations, it matters which strategy space is being explored. A basic question is whether a strategy taken from the space always has a best reply within the space. Lev\'{\i}nsk\'{y} et al~\citep{SI:Levinsky:IJGT:2020} answer this question for factored strategies, which choose their actions not on the basis of the entire history, but on the basis of a function $\varphi$ of the history, called a factor. Memory-$n$ strategies are the special case where $\varphi$ records the outcomes of the last $n$ rounds. They show that if $\varphi$ is recursive, which means that the stored information can be updated incrementally, then every $\varphi$-factored strategy admits a $\varphi$-factored best reply, and the best reply property holds in all subgames. Building on this result, LaPorte et al~\citep{SI:LaPorte:PNAS:2026} formalize two notions of completeness. A space is best-reply-complete if every strategy in the space has a best reply within the space; it is payoff-complete if any strategy playing against an opponent from the space can be replaced by a payoff-equivalent strategy in the space. They give conditions for these properties and construct complete spaces for simple games. In the present paper we explore the spaces of M1 and of M2 strategies.  We find that the additional strategies, which become available for M2, allow evolution to escape the Hold-trap in SD games.\\

\noindent
\textbf{Payoff from a limited number of interactions.}
Classical models of evolutionary game dynamics assume that each individual interacts with all other members of the population, or with a representative sample of it, so that payoffs are deterministic and given by their expected values. Some studies relax this assumption. Traulsen et al~\citep{SI:Traulsen:JTB:2006} let each individual have a randomly distributed number of interactions, so that two individuals who use the same strategy can obtain different payoffs. They find that the resulting payoff stochasticity reduces the effective intensity of selection, or in their terminology increases the temperature of selection. Woelfing and Traulsen~\citep{SI:Woelfing:JTB:2009} compare stochastic sampling of interaction partners with deterministic payoff assignment. Hilbe~\citep{SI:Hilbe:BMB:2011} studies an infinite population in which interactions and reproduction occur in random groups of fixed size, and shows that the resulting dynamics is the replicator equation for a slightly modified payoff matrix, which reproduces results for finite populations such as the one-third law. In direct reciprocity, the assumption that a single pair interacts per round is used by Schmid et al~\citep{SI:Schmid:NHB:2021} and by H\"ubner et al~\citep{SI:Hubner:PNASnexus:2025}. Our model belongs to this class: each individual chooses $m$ co-players, and we typically use $m=1$. This choice makes the running time of the algorithm roughly linear in population size, which is what allows us to study populations of up to $N=10^5$. It also means that payoffs are stochastic in the above sense. SI Figure 6a shows that increasing $m$ from 1 to 100 has essentially no effect on our results.\\

\noindent Most models of direct reciprocity assume that individuals compare expected payoffs, averaged over all interactions in the population. Glynatsi et al~\citep{SI:Glynatsi:PRSB:2024} relax this assumption and let individuals update their strategies based on the payoffs of their most recent interactions. They find that cooperation still evolves, but that individuals adopt less generous strategies. In the extreme case, when only the very last interaction is taken into account, the maximum generosity is bounded by $1/2$, independently of the benefit-to-cost ratio, while generosity is unbounded if average payoffs are used. Our model uses realized rather than expected payoffs: each individual plays $m$ interactions, typically $m=1$, and the total payoff obtained in those interactions is used for pairwise comparison. Our results therefore hold in a setting where individuals act on very limited payoff information.\\

\noindent

\noindent
\textbf{The role of longer memory.}
Some research has investigated evolutionary dynamics among strategies with longer memory~\citep{SI:Hauert:ProcB:1997,SI:hauert2002simple,SI:stewart:scirep:2016,SI:hilbe:pnas:2017,SI:Do:JTB:2017,SI:Murase:ScientificReports:2020,SI:li2022evolution,SI:Murase:PLoSCompBio:2023a,SI:Glynatsi:PNAS:2024}.
These studies suggest that longer memory facilitates evolution of cooperation.
An equilibrium analysis shows that longer memory allows for strategies that can sustain full cooperation even if $b/c$ becomes small \citep{SI:hilbe:pnas:2017}. For $b/c = 1.2$, the suggested strategy needs to
memorize the outcome of the last five rounds. It is, however, unclear
if evolution would  find such strategies. Already for memory-3, a systematic exploration of the strategy space is  infeasible due to the huge number of possible strategies
\citep{SI:hilbe:pnas:2017}. As a result, it is not known to which extent higher
memory can effectively promote the evolution of cooperation for small $b/c$
ratios.\\

\subsection{Related literature on the role of mutation in game dynamics}\label{SI:section:mutation}

\noindent
\textbf{Game dynamics without mutation.}
Frequency dependent selection among a fixed number of strategies~\citep{SI:Brandt:JTB:2006,SI:Hilbe:BMB:2011}.  
can be studied with the classical replicator equation
\citep{SI:taylor:MB:1978}, which describes a
deterministic evolutionary process in infinitely large populations in the
absence of mutation. The resulting dynamics depend on which
strategies are being considered. For example, between
ALLD and GTFT, replicator dynamics find bistable
competition \citep{SI:nunez2016evolution}. For other strategies, there can be
cycles \citep{SI:nowak1989game} or coexistence  \citep{SI:Grujic:JTB:2012}.\\

\noindent Researchers have also
explored direct reciprocity in structured populations. One common assumption is
that individuals are arranged on a regular lattice, and that they only interact
with their immediate neighbors. The respective simulations often
assume that mutation does not introduce new strategies
\citep{SI:nowak1992spatialgames,SI:szabo2000spatial,SI:szolnoki2014defection,SI:Szolnoki:PRE:2014}.\\

\noindent
\textbf{Game dynamics with rare mutations.} 
Many studies of direct reciprocity assume vanishingly small
mutation rates \cite{SI:Kurokawa:PRSB:2009,SI:imhof:PRSB:2010,SI:van-segbroeck:PRL:2012,SI:Pinheiro:PLoS:2014}. 
In the rare-mutation limit, populations are almost always homogeneous, and evolutionary dynamics can be characterized in terms of fixation probabilities of single mutants~\cite{SI:nowak:Nature:2004,SI:fudenberg:JET:2006,SI:wu:JMB:2012}.\\

\noindent The assumption of vanishingly small mutation rates is made
in studies of adaptive dynamics \citep{SI:nowak:AAM:1990,SI:geritz1998evolutionarily}. 
This framework has been used to explore the space of reactive and of M1 strategies
\citep{SI:nowak:AAM:1990,SI:stewart:games:2015,SI:reiter2018crosstalk,SI:LaPorte:PLoSCB:2023}.
LaPorte et al~\citep{SI:LaPorte:PLoSCB:2023} derive the adaptive dynamics for the full space of M1 strategies in the repeated donation game. They show that the four-dimensional space of M1 strategies contains an invariant three-dimensional subspace, which consists of the counting strategies. Counting strategies react only to the number of cooperators in the previous round, and therefore satisfy $p_2=p_3$. They also describe a symmetry between orbits forward and backward in time, which is specific to the donation game. These results allow them to characterize the adaptive dynamics among M1 counting strategies completely, and among all M1 strategies in part. Adaptive dynamics assumes that mutants arise rarely and differ only slightly from the resident. The present paper explores the opposite regime: mutants are common and they are drawn from the entire strategy space.\\

\noindent Imhof and Nowak \citep{SI:imhof:PRSB:2010} study evolutionary dynamics in the rare-mutation limit with mutants drawn from the entire space. They exploit the fact that the fixation probability of a single mutant in a homogeneous resident population can be computed explicitly \citep{SI:nowak:Nature:2004}. Since then, the rare-mutation assumption has been widely used to explore direct reciprocity in populations of finite size
\citep{SI:Hilbe:PNAS:2013,SI:Stewart:PNAS:2013,SI:Stewart:PNAS:2014,SI:%
stewart2016evolutionary,SI:Baek:SciRep:2016,SI:donahue2020evolving,SI:%
Schmid:NHB:2021,SI:park2022cooperation,SI:willensdorfer2005mutation,SI:%
Kurokawa:PRSB:2009,SI:van-segbroeck:PRL:2012,SI:Pinheiro:PLoS:2014}.\\

\noindent
\textbf{Studies with fixed mutation rate.} 
Some of the early studies of direct reciprocity consider simulations with a strictly positive mutation rate
\citep{SI:Nowak:Nature:1992,SI:Nowak:Nature:1993,SI:Hauert:ProcB:1997}. In many of
those studies, the mutation rate is not varied.\\

\noindent
\textbf{Studies that vary mutation rate without finding a positive
effect.} In some cases, researchers have varied the mutation rate without
discovering a positive effect of intermediate mutation rates
\citep{SI:donahue2020evolving,SI:Schmid:NHB:2021,SI:park2022cooperation,SI:%
willensdorfer2005mutation}. Since the respective studies did not focus on the
impact of mutations, they did not explore the parameter space systematically.
In many cases, simulations have been run for parameter values that naturally
favor cooperation even if mutations are rare.\\

\noindent
\textbf{Biased mutations.}
Almost all of the above studies draw mutant strategies uniformly at random from the strategy space. A few studies depart from this convention. Schmid et al~\citep{SI:Schmid:PLoSCB:2022} compare players who use unconditional, reactive and M1 strategies. Among the sampling schemes they consider is one they call same-complexity sampling with biasing: a player draws two random numbers and adopts the more extreme of the two, which is the one closer to the boundary of the interval $[0,1]$. This is the same device we use here, although for a different purpose. They introduce the bias so that players with less memory can find extreme strategies as easily as players with more memory. We use it to speed up the discovery of effective strategies in the M1 and M2 spaces. Garc\'ia and Traulsen~\citep{SI:Garcia:PLoSONE:2012} point out that uniform mutation is rarely justified, because not every strategy is equally likely to arise from a given resident. Comparing uniform mutation with bitwise mutation among deterministic strategies with one round of memory, they find that the more restricted mutations shift the outcome toward less cooperation. Stewart and Plotkin~\citep{SI:stewart:games:2015} compare local mutations, which generate mutants in the vicinity of the resident, with non-local mutations, which are drawn from the whole M1 space. They find that local mutations admit a wider range of evolutionarily robust outcomes, but make mutual cooperation harder to evolve. It is worth noting that in both of these studies mutations that are more local lead to less cooperation, whereas our mutations are non-local and biased toward the boundary, and lead to more cooperation. The parameter $\alpha$ controls the strength of the bias: for $\alpha=0$ we recover uniform mutation, and for $\alpha=1/2$ we obtain binary strategies.\\

\noindent
\textbf{The boundary of the strategy space.}
There is an analytical reason why the boundary deserves attention. LaPorte et al~\citep{SI:LaPorte:PLoSCB:2023} study the adaptive dynamics of M1 strategies in the repeated donation game. They define the boundary of the strategy cube as the set of points with exactly one component $p_i \in \{0,1\}$, which is a union of eight three-dimensional faces, and they characterize the saturated points on those faces. On the face $p_1=1$, which means cooperation after mutual cooperation, mutual cooperation is absorbing and the payoff is $b-c$, irrespective of the other three components. They also determine numerically where trajectories end. For $c/b=0.1$ and a grid of $9^4=6561$ initial conditions, 1835 trajectories converge to full cooperation, 1375 to full defection, 2964 to other points on the boundary and only 387 to interior critical points, which are equalizers. Hence, more than $90\%$ of all trajectories terminate on the boundary. This supports our choice of a mutational process that is biased toward the boundary.\\

\noindent
\textbf{Deterministic strategies.}
For $\alpha=1/2$ our mutational process generates only strategies whose components are close to 0 or 1, which we call binary strategies. Deterministic strategy spaces have been studied before. Baek et al~\citep{SI:Baek:SciRep:2016} compare deterministic and stochastic strategies, both for reactive and for M1 players. Murase and Baek~\citep{SI:Murase:ScientificReports:2020,SI:Murase:PLoSCompBio:2023a} study deterministic strategies with a memory of up to three rounds. Tkadlec et al~\citep{SI:Tkadlec:PNAS:2023} include a simulation among all deterministic M2 strategies in their Supplementary Information (see below).\\

\noindent
\textbf{Effect of mutations on evolutionary stability.} In an important early work, it was shown that if mutants are allowed
to be heterogeneous, no pure strategy is evolutionarily stable \citep{SI:boyd:Nature:1987}. For any resident population, it is possible to construct mixtures of mutant strategies that are favored in the respective environment. This line of research has been extended in subsequent studies
\citep{SI:lorberbaum1994nostrategy,SI:lorberbaum2002really,SI:vanveelen2012direct,SI:%
garcia:jet:2016,SI:Garcia:Frontiers:2018}. Importantly, however, this instability
is symmetric: it affects cooperative and non-cooperative equilibria alike.\\

\noindent\textbf{Previous studies on the effect of mutation on cooperation more
generally.} Some studies find a positive effect of mutations in other games.
Traulsen et al \citep{SI:traulsen2009exploration} explore the evolution of peer punishment. In their model,
players first decide whether or not to contribute to a public good. In a second stage, players decide whether to punish non-contributors. Their baseline model considers three strategies: defectors, who neither contribute nor punish;
cooperators who contribute but do not punish; and punishers who both contribute and punish. They find, if mutations are rare, most players defect, but if mutation are common, cooperators prevail.\\

\noindent Ramírez et al \citep{SI:ramirez2022diversity}
explore evolutionary dynamics of the traveler's dilemma. In this game, two
players  choose an integer within an interval $[L,U]$. A player's payoff
is given by the lower of the two integers. In addition, if the players' chosen
integers are different, the player with the lower integer obtains an additional
reward $R$. Backward induction suggests the game has a unique,
inefficient equilibrium: both players choose $L$. The authors
find that players learn to choose large numbers, if the reward is 
small and mutations are common.\\

\noindent
 McNamarra et al \citep{SI:mcnamara2004variation} study a finitely
repeated PD. Their model is as follows. If both players
cooperate in all rounds, the total number of rounds is $N$. Otherwise,
if any of the players defects before round $N$, the game stops after that
defection. The rules of the game and the parameter $N$ is known to the players.
Again, the game can be solved by backward induction. In the unique,
inefficient equilibrium, both players defect immediately. For their
evolutionary analysis, the authors consider strategies that are given by a
threshold $n$. A player with threshold $n$ cooperates until round $n$ and
defects in round $n{+}1$, unless the game is stopped before. Evolutionary
simulations suggest that players choose a small threshold when mutations are
rare, consistent with backward induction. Once mutations are common, 
larger thresholds are favored.\\

\noindent The mechanism that leads to cooperation in the above studies is different from
ours. In the above studies
\citep{SI:traulsen2009exploration,SI:ramirez2022diversity,SI:mcnamara2004variation},
full cooperation is generally unstable. The role of mutations is to
introduce strategies that give a payoff advantage to
cooperators. In contrast, in our setup, cooperation is generally stable. While
there are equilibria in which everyone defects, there are also equilibria in
which everyone cooperates. The role of mutations, in our study, is to help
evolution to find these cooperative equilibria more efficiently, and to make
them more robust against neutral invasions.\\

\noindent\textbf{Work on the importance of variation and diversity.} 
In group selection, cooperation can evolve if different groups compete \citep{SI:boyd1982cultural}. 
Group selection is effective if there is variation between groups but not within groups. 
Once individual groups become too heterogeneous, 
group selection fails to select for cooperation \citep{SI:traulsen2006evolution}.
The interplay of diversity and cooperation was also studied
in the context of spatial adaptive dynamics~\citep{SI:hauert2021spatial}.\\

\noindent
\textbf{Tkadlec et al.}~\cite{SI:Tkadlec:PNAS:2023}
use a frequency dependent Wright Fisher process with pairwise comparison and deterministic payoff evaluation.
In their simulations all individuals interact with all others. 
Therefore, simulations were limited to small population sizes (of $N=100$ and 200).
Moreover, they use a mutational process without bias, and therefore much longer simulation runs were needed.
They only study the donation game.
Their main result is that mutation enhances cooperation.
In their Supplementary Information, Tkadlec et al also study evolutionary dynamics among binary M2 strategies in the donation game, varying the cost of cooperation and the mutation rate, for population size $N=100$ and intensity of selection $\beta=10$. They restrict themselves to binary M2 strategies, because they consider the space of stochastic M2 strategies difficult to explore exhaustively with simulations. They find that cooperation evolves for small costs; that for intermediate costs it evolves only for intermediate mutation rates; and that for high costs cooperation hardly evolves at all. Our findings agree with theirs for that population size. For the donation game with $c=20/21\approx0.95$, which is the highest cost they study, we obtain a median payoff of 0.54 for $N=100$, but 0.97 for $N=10^3$ and 0.99 for $N=10^4$ (SI Figure 7). The main point of the present paper is that cooperation is restored once the population is large.\\ 

\subsection{Novelty of this paper}\label{SI:section:novelty}

The novelty of this paper is the observation that essentially all two-person social dilemmas can be resolved to maximum payoff by simple and natural processes of evolutionary dynamics that uses pairwise comparison. The crucial insight is to consider large population size ($N=10^3$ or more) and strong (but not too strong) intensity of selection. Then there is a broad window of mutation rates, that enable the population to find maximum payoff in PD, SD, SH and HY games, even for extremely unfavorable payoff values, and also for situations where alternating gives a higher payoff than complete cooperation.  M1 strategies suffice to resolve all PD, SH and HY games, while M2 strategies (and also binary M2 strategies) resolve all four games. When considering continuous strategy spaces, it is important that the mutational process is biased toward the boundaries.
Equilibrium analysis establishes that strategies which sustain the socially optimal outcome exist in every social dilemma~\citep{SI:Hubner:PNASnexus:2025}. Our contribution is dynamical: we identify the regime of population size, mutation rate and intensity of selection in which a simple evolutionary process actually finds them.\\

\section{Donation game}

The donation game is the benchmark for testing evolutionary dynamics of cooperation. A cooperator pays a cost, $c$, for the co-player to receive a benefit, $b$. A defector pays no cost and confers no benefit. The payoff matrix per round is given by 
\begin{equation} \label{SI:E1}
    \bordermatrix { & C & D \cr
                    C & b-c & -c \cr
                    D &b & 0}                    
\end{equation}
The parameters obey $b>c>0$. Without loss of generality we set $b=1$ and vary $c$ between 0 and 1. In the other parameterization used in this paper, we obtain donation games along the half line $u+v=0$ with $v>0$. For any such game, benefit and cost are $b=1$ and $c=v/(1+v)$.\\

\noindent \textbf{Efficiency is the cooperation rate.} For the donation game the scaled payoff has a simple interpretation. Consider two players of a repeated game and let $\gamma_1$ and $\gamma_2$ be the frequencies with which they cooperate. Their payoffs per round are $\pi_1 = b\gamma_2 - c\gamma_1$ and $\pi_2 = b\gamma_1 - c\gamma_2$, and therefore
\begin{equation} \label{SI:eq:pairsum}
    \pi_1 + \pi_2 = (b-c)\,(\gamma_1 + \gamma_2).
\end{equation}
In each generation there are $Nm$ games and hence $2Nm$ player roles, and every game contributes to the payoff of both of its players. Summing~\eqref{SI:eq:pairsum} over all games and dividing by $2Nm$ shows that the average payoff per interaction is $\bar\pi = (b-c)\,\bar\gamma$, where $\bar\gamma$ is the average cooperation rate of the population. For the donation game $\pi_{\max}=b-c$ and $\pi_{\min}=0$, and hence
\begin{equation} \label{SI:eq:effgamma}
    \bar\pi_s = \frac{\bar\pi}{b-c} = \bar\gamma .
\end{equation}
Efficiency and cooperation rate are therefore the same quantity - exactly, and not only in the limit of long runs. Every efficiency value that we report for a donation game can be read directly as a percentage of cooperation.\\

\noindent \textbf{The limit of maximum cost.} Equation~\eqref{SI:eq:pairsum} also shows why the donation game becomes hard as the cost approaches the benefit. With $b=1$ the payoffs of a pair sum to $(1-c)(\gamma_1+\gamma_2)$, which vanishes as $c \to 1$: in that limit the game is zero-sum and cooperation is a pure transfer from one player to the other. Moreover, for any two strategies,
\begin{equation} \label{SI:eq:pairdiff}
    \pi_1 - \pi_2 = (b+c)\,(\gamma_2 - \gamma_1),
\end{equation}
so the player who cooperates less always obtains the higher payoff, and this difference stays of order one however close $c$ is to $b$. At $c=b$ the payoff for mutual cooperation equals the payoff for mutual defection, and no strategy - with a memory of any length - can make cooperation a best reply. Thus $b/c=1$ is not a soft boundary but the exact point at which the surplus that reciprocity could enforce vanishes.\\

\noindent For $c<1$ the surplus per round is $1-c>0$, but the temptation to defect against a cooperator is $T-R=c$, which is of order one. The advantage that a cooperative community holds over a defector is therefore of order $1-c$, while the gradient favoring exploitation is of order one. The intensity of selection multiplies both and cannot alter their ratio; what remains is population size and mutation rate. This is why the process needs larger populations as $c$ approaches one: for $c=0.99$ cooperation fails entirely below $N \approx 2.5\times10^3$ and then rises monotonically with $N$ (SI Figure 4b).\\

\noindent SI Figures 1-3 show evolutionary simulations of the infinitely repeated donation game with $c=0.95$ and $b=1$ using M1 strategies. We study three population sizes, $N=100, 500, 1000$, and three intensities of selection, $\beta=1 , 10, 100$. The mutation rate varies between $\mu=10^{-5}$ and $10^{-1}$.  For each triple $(N,\beta,\mu)$ we average over 10 runs. Each run is conducted for $10^8$ generations. Measurement starts after a burn in of $10^6$ generations. We use mutational bias, $\alpha=1/3$ and noise floor, $\epsilon=10^{-4}$, with $p_{min}=\epsilon$ and $p_{max}=1-\epsilon$.\\

\noindent The state space of the stochastic process is high dimensional and continuous. For measuring oscillations between cooperative and defective populations, we divide the state space into three regions: (i) a cooperative state, (ii) an intermediate state, and (iii) a defective state.  For judging transitions between those states we use hysteresis: if the scaled average payoff, $\bar\pi_s$, climbs above 0.9 we enter the cooperative state; if $\bar\pi_s$ falls below 0.7 we leave the cooperative state;  if $\bar\pi_s$ falls below 0.1 we enter the defective state; if $\bar\pi_s$ climbs above 0.3 we leave the defective state; at other times we are in the intermediate state.  The hysteresis avoids counting small fluctuations when evaluating mean dwell times in each state.\\

\noindent SI Figure 1 shows the fraction of times the process spends in the cooperative state. For $N=100$ and $\beta=100$ there is maximum near mutation rate $\mu=0.014$. For $N=1000$ and $\beta=100$ there is a broad peak between mutation rates 0.0003 and 0.04 where the process is in the cooperative state more than 90\% of the time.\\

\noindent SI Figure 2  shows the fraction of time the population spends in the
cooperative (blue), intermediate (yellow), and defective (red) states as a
function of the mutation rate $\mu$, for selection strengths
$\beta = 0.1$, 10, 100, 1000, 10000 (rows) and population sizes
$N = 100, 500, 1000$ (columns). At weak selection
($\beta = 0.1$) cooperation is essentially absent for all $N$: the
population sits in the defective state at low $\mu$ and shifts to the
intermediate state at high $\mu$. As selection strengthens to $\beta = 10$ and
$\beta = 100$ the cooperative state comes to dominate over an intermediate band
of mutation rates, forming a hump that broadens and deepens with both $N$ and
$\beta$; for $N = 1000$, $\beta = 100$ the population cooperates almost all of
the time from $\mu \approx 10^{-4}$ up to the high-$\mu$ collapse.
Stronger selection does not, however, keep improving cooperation---the
dependence on $\beta$ is non-monotonic. At very strong selection
($\beta = 1000$ and $10000$) the cooperative window
narrows and shifts to higher $\mu$: for $N = 1000$ the onset of majority
cooperation moves from $\mu \approx 10^{-4}$ at $\beta = 100$ to
$\approx 8\times10^{-4}$ at $\beta = 1000$ and $\approx 6\times10^{-3}$ at
$\beta = 10000$, and the smallest population is hit hardest---at $N = 100$,
$\beta = 10000$ cooperation essentially fails (peak occupancy $\approx 0.15$),
while the larger populations still reach near-complete cooperation but only
within a thin high-$\mu$ band. This reflects the two-sided role of selection
intensity: stronger selection deepens both basins, so the cooperative state is
more stable once reached but the defective state is correspondingly harder to
escape by drift and requires a larger mutational supply (higher $\mu$) to
nucleate cooperation. Cooperation thus fails at low $\mu$---where the supply is
insufficient, a boundary that moves to higher $\mu$ as $\beta$ increases---and
again at high $\mu$, where mutational load erodes the cooperative strategy, and
is maximised at intermediate selection strength and intermediate mutation rate
in the larger populations.\\

\noindent SI Figure 3 shows the mean dwell time---the average number of
consecutive generations the population remains in a state before leaving
it---for the cooperative, intermediate, and defective states (one panel each)
as a function of the mutation rate $\mu$, for population sizes
$N = 100, 500, 1000$ and $\beta = 10$. Cooperative spells are longest in the larger populations, reaching
$\sim\!5\times10^{4}$ generations for $N = 1000$.\\

\noindent The cooperative panel reveals the decisive difference between small and large
populations. For $N = 100$ the cooperative dwell time declines
monotonically with $\mu$: mutation only ever shortens cooperative
episodes, so the longest spells occur in the low-mutation limit. For $N = 500$ and
$N = 1000$ the dependence is instead non-monotonic: the dwell time first dips
and then recovers to an interior maximum before falling away at high $\mu$.
For $N = 500$ that recovery peak, $\sim\!1.5\times10^{4}$ generations near
$\mu \approx 5\times10^{-4}$, does not quite regain the low-mutation value.
Only for $N = 1000$ does mutation lengthen cooperative episodes outright,
reaching $\sim\!5\times10^{4}$ generations near $\mu \approx 3\times10^{-4}$,
roughly $2.5$ times its value in the low-mutation limit. In a
sufficiently large population a moderate supply of mutants therefore
prolongs cooperation whereas in a small population the
same mutations merely destabilize the shallow cooperative basin and shorten its
lifetime. This reversal, with mutation stabilizing cooperation in large
populations but destabilizing it in small ones, is a central reason why larger
populations promote cooperation.\\

\noindent The defective state shows a u-shape behavior: spells last of order
$10^{5}$ generations at low $\mu$ (from $8\times10^{4}$ for $N = 1000$ up to
$2.7\times10^{5}$ for $N = 100$), where recovery of cooperation is
supply-limited (and faster in larger populations), fall to a minimum of only a
few hundred generations at intermediate $\mu \approx 5\times10^{-3}$, and
lengthen again as $\mu$ increases further. The intermediate state is short-lived
(tens to a few hundred generations) across most of the range, but its dwell time
rises steeply at high $\mu$ for the larger populations, as the system becomes
trapped in the intermediate state near the error threshold.\\ 

\noindent SI Figure 4 shows the average payoff versus intensity of selection, $\beta$,
and population size, $N$, for two cost values, $c=0.95$ and $c=0.99$ using M1 strategies. 
The dependence on $\beta$ has a broad peak: cooperation is
low under near-neutral drift ($\beta\lesssim1$, with a shallow minimum at
$\beta\approx1$), rises steeply to
a wide plateau at intermediate selection ($\beta\approx10^{2}$--$10^{3}$;
payoff $\approx0.97$ for $c=0.95$, $\approx0.93$ for $c=0.99$), and then
collapses once selection becomes too strong. The near-deterministic
imitate-the-fitter regime traps the population in defection. 
In contrast, for population size, $N$, the dependence appears to be monotone: payoff rises with $N$ and
saturates near unity, with no upper limit up to $N=10^{5}$. In both panels the
harder game ($c=0.99$) needs more of the ordering force: its cooperative
window in $\beta$ is narrower and shifted to stronger selection, and in $N$ it
fails entirely below $N\approx2.5\times10^{3}$ before rising to
$\approx0.91$ at $N=10^{5}$; its persistently large run-to-run scatter reflects
the metastable, bistable dynamics at the near-neutral benefit margin
$b-c=0.01$. Together the panels show that cooperation requires an
intermediate (but high) intensity of selection but benefits monotonically from larger
populations.\\

\noindent Figure 5 of the main text shows average scaled payoff as a
function of the cost $c$ for three parameter choices:
$N = 10^{3},\ \beta = 100$ (blue), $N = 10^{4},\ \beta = 10$ (yellow), and
$N = 10^{4},\ \beta = 100$ (red). Bars give the mean over $30$ replicate runs,
the dots the individual runs, and the error bars $\pm 1$ standard deviation. At
costs $c = 0.95$ and 0.99, all three choices sustain near-complete cooperation.
The payoff declines as the cost approaches the
benefit ($c \to 1$, with $b = 1$). The very large population ($N = 10^{4}$,
yellow and red) is better at maintaining cooperation than the
smaller one ($N = 10^{3}$, blue), the gap widening as $c$ increases; among them the
more strongly selected population ($\beta = 100$, red) is consistently best.
Population size is thus the dominant factor in sustaining cooperation at high
cost, with selection strength providing a further edge. Cooperation 
collapses sharply between $c = 0.997$ and $c = 0.998$, where the payoff falls
essentially to zero for all three choices (only the red curve retains a small,
highly variable residue). The run-to-run scatter is largest at the marginal
costs $c \approx 0.995$--$0.997$ (standard deviations up to $\sim\!0.15$),
reflecting the bistable, metastable regime in which an individual population
either sustains or loses cooperation, whereas at lower and higher cost the
outcome is nearly deterministic.\\

\section{Robustness toward parameter variation}

\noindent Now we describe additional simulations, which evaluate the effect of variation in crucial parameters.  Simulations start with random M1 or M2 strategies and run for $10^6$ generations. Average payoff is sampled in the second half of the run. Unless otherwise stated, the standard parameters are:  population size $N=1000$, mutation rate $\mu=10^{-2}$, intensity of selection $\beta=10$, interaction partners $m=1$, noise floor $\epsilon=10^{-4}$, mutational bias $\alpha=1/3$.\\  

\noindent  All reported payoffs are scaled relative to the maximum and minimum payoff that can be achieved in a particular game. We use $\bar\pi_s = (\bar \pi -\pi_{min})/ (\pi_{\max}-\pi_{\min})$ where $\bar \pi$ is the average payoff in a population.
For memory-n strategies (with $n \geq 1$), the maximum and minimum per round payoffs that can be achieved 
in the infinitely repeated game are given by $\pi_{\max}= \max \{ 1, (1+v+u)/2\}$ and 
$\pi_{\min}= \min \{ 0, (1+v+u)/2\}$.\\ 

\subsection{Varying the bias in the mutational process, $\alpha$}

\noindent  The parameter, $\alpha \in [0,1/2]$, determines the mutational bias toward the boundary. The random draw for each $p_i$ value of a mutant strategy has three outcomes: (i) with probability $\alpha$ it is $p_{min}=\epsilon$; (ii) with probability $1-2\alpha$ it is a uniformly distributed value in the interval $(p_{min},p_{max})$; (iii) with probability $\alpha$ it is $p_{max}=1-\epsilon$. For $\alpha=0$ there is no bias toward the boundary, and mutants are drawn uniformly at random from the open interval $(0,1)$.  For $\alpha=0.5$, there is maximum bias toward the boundary and the strategy space is reduced to the binary choice between $p_{min}$ and $p_{max}$. For $0<\alpha<0.5$, mutants are drawn from the closed interval $[p_{min}, p_{max}]$ with a weight on the boundary.\\

\noindent  SI Figure 5 shows the effect of varying $\alpha$.  For  $\alpha=0$, many games remain unresolved; $\alpha=0.1$ helps M1 strategies to resolve many games outside of the Hold-trap, but is rather inefficient for M2; $\alpha=0.2, 0.3, 0.4$ work well for both M1 and M2; using only binary strategies ($\alpha=0.5$) succeeds for M2 but not for M1.\\

\subsection{Varying the number of interaction partners, $m$}

Each individual chooses $m$ consecutive co-players at random. Self-interactions are avoided. Therefore each individual plays at least $m$ games, the average number of games per individual is $2m$, but the actual number of games per individual varies. For each individual we consider the total payoff sum without normalizing by number of interactions. SI Figure 6a shows there is essentially no variation in outcome as $m$ changes between 1 and 100. Being able to simulate the process for $m=1$ is a dramatic speed up in computation time for large populations.\\

\subsection{Varying the intensity of selection, $\beta$} 

For pairwise comparison we use a Fermi function with parameter $\beta$. SI Figure 6b shows that high efficiency is obtained for a broad range of $\beta$ values. For $\beta=1000$ we lose efficiency in some games.\\

\subsection{Varying the mutation rate, $\mu$}

During updating, an individual chooses a new mutant with probability $\mu$ or a random role model for pairwise comparison with probability $1-\mu$. SI Figure 6c shows that high payoffs are achieved for a broad range of intermediate mutation rates.\\ 

\subsection{Varying the minimum noise level, $\epsilon$}

Mutants are drawn with $p$ values that are at least $p_{min}=\epsilon$ and at most $p_{max}=1-\epsilon$. SI Figure 6d shows that efficiency is lost in some games when $\epsilon$ becomes too large, around $10^{-2}$.\\

\subsection{Memory-2 strategies resolve games with extreme payoffs}

So far in all simulations of the four games we varied the payoff parameters, $u$ and $v$, from $-8$ to $+8$. Now we zoom out to consider a wider range, $-40$ to $+40$. For example, in the top left corner of this game space, there is a Prisoners Dilemma with payoff matrix
\begin{equation} \label{SI:E2}
    \bordermatrix { & C & D \cr
                    C & 1 & -40 \cr
                    D &41 & 0}                    
\end{equation}

\noindent SI Figure 7 shows the ability of M2 strategies to resolve games with such extreme payoff values. Population sizes $N=100, 1000, 10000$ are shown. Each point in the heat map is the median over 10 independent runs.\\

\section{Finitely repeated games} 

So far we have studied infinitely repeated games. Now we examine if our findings are robust for finitely repeated games. After each round the game ends with probability, $d$. M1 strategies for finitely repeated games are given by 5 parameters, $(p_0,p_1,p_2,p_3,p_4)$. The parameters $p_1,p_2,p_3,p_4$ are the same as before. The additional parameter $p_0$ is the probability to cooperate in the first round. SI Figure 8 shows that the results are essentially unchanged for finitely repeated games with $d=0.001$.\\ 

\section{A process with asynchronous updating} 

Until now we have assumed synchronous updating: all individuals update at the same time. Now we investigate a process with asynchronous updating. Initially all individuals adopt random strategies. Then all individual interacts in some games. The initial payoff of an individual is the average payoff received in those interactions. Then at each time step one individual is chosen at random for updating. With probability $\mu$ the individual updates by mutation. With probability $1-\mu$ the individual updates by pairwise comparison with a randomly chosen role model. Independent of updating, $m$ pairs of individuals are chosen to interact in a game. After each game the new payoff of an individual is the average of the existing payoff and the payoff achieved in the most recent interaction. The process is repeated over many time steps and the scaled average payoff is recorded in the second half of the run. SI Figure 9 shows simulations of this process for M2 strategies. For population size $N=100$ some games remain unresolved. For $N=1000$ almost all games are resolved.\\ 

\section{Expected payoff against a random M1 population}

Here we use the general payoff matrix $R,S,T,P$. A player receives $R$ for mutual cooperation,
$P$ for mutual defection, $S$ when cooperating against a defector, and $T$
when defecting against a cooperator. A memory-1 (M1) strategy is a vector
$p=(p_1,p_2,p_3,p_4)$, where $p_i$ is the probability of cooperating after the
outcome $CC,CD,DC,DD$, listed with the focal player's own move first. 
Against an opponent $q$ the pair of moves performs a
Markov chain on the four outcomes. Since the opponent labels the same outcomes
from the opposite side, $q$ enters through $\tilde q=(q_1,q_3,q_2,q_4)$, and
the transition matrix is
\[
  M_{i\,\cdot}\;=\;\bigl(p_i\tilde q_i,\;
                         p_i(1-\tilde q_i),\;
                         (1-p_i)\tilde q_i,\;
                         (1-p_i)(1-\tilde q_i)\bigr)
\]
We consider $\eps\le p_i\le 1-\eps$. Since every entry of $M$ is at least $\eps^{2}>0$, there is a unique stationary
distribution $v(p,q)=(v_{CC},v_{CD},v_{DC},v_{DD})$. The long-run payoff is
the linear functional
\begin{equation}
  s(p,q)\;=\;v_{CC}\,R+v_{CD}\,S+v_{DC}\,T+v_{DD}\,P 
  \label{SI:eq:payoff}
\end{equation}
Let $\Ran$ denote an opponent drawn uniformly from the strategy space
$\Omega_\eps=[\eps,1-\eps]^4$, and write
\[
  A(p)=\Ex_q[v_{CC}],\quad B(p)=\Ex_q[v_{CD}],\quad
  C(p)=\Ex_q[v_{DC}],\quad D(p)=\Ex_q[v_{DD}],
\]
so that $A+B+C+D=1$ and, because the stationary distribution does not depend on
the payoff values,
\begin{equation}
  E(p,\Ran)\;=\;A(p)\,R+B(p)\,S+C(p)\,T+D(p)\,P
  \label{SI:eq:affine}
\end{equation}
is \emph{exactly linear} in $(R,S,T,P)$, with coefficients that are properties
of the strategy alone.

\paragraph{A symmetry.}
Exchanging the labels $C$ and $D$ for both players permutes the outcomes by
\[
  \pi:\;CC\leftrightarrow DD,\qquad CD\leftrightarrow DC ,
\]
and turns the strategy $p$ into its \emph{conjugate}
\begin{equation}
  p^{\,c}_s \;=\; 1-p_{\pi(s)},
  \qquad\text{i.e.}\qquad
  p^{\,c}=(1-p_4,\;1-p_3,\;1-p_2,\;1-p_1).
  \label{SI:eq:conj}
\end{equation}
Conjugation commutes with the role swap $q\mapsto\tilde q$, so relabelling the
states by $\pi$ carries the chain of $(p,q)$ exactly onto the chain of
$(p^{\,c},q^{\,c})$, i.e.\ $M^{p^c,q^c}_{\pi(i)\pi(j)}=M^{p,q}_{ij}$, and
therefore
\begin{equation}
  v_s(p^{\,c},q^{\,c})\;=\;v_{\pi(s)}(p,q)\qquad\text{for all }s .
  \label{SI:eq:lemma}
\end{equation}
Now note that both strategies of interest are \emph{self-conjugate}, since
\eqref{SI:eq:conj} reverses the vector and complements each entry, and each of
them is invariant under exactly that operation:
\[
  \SWAP=(\tfrac12,\eps,1-\eps,\tfrac12)
       \;\stackrel{c}{\longmapsto}\;(\tfrac12,\eps,1-\eps,\tfrac12),
  \qquad
  \HOLD=(\tfrac12,1-\eps,\eps,\tfrac12)
       \;\stackrel{c}{\longmapsto}\;(\tfrac12,1-\eps,\eps,\tfrac12).
\]
The cube $\Omega_\eps$ is invariant under $q\mapsto q^{\,c}$ and the map
preserves Lebesgue measure, so substituting $q\mapsto q^{\,c}$ in the average
and applying \eqref{SI:eq:lemma} with $p=p^{\,c}$ gives, for any self-conjugate
$p$,
\begin{equation}
  A(p)=D(p),\qquad B(p)=C(p),
  \qquad\text{hence}\qquad A(p)+B(p)=\tfrac12 .
  \label{SI:eq:AD}
\end{equation}

\paragraph{Consequence.}
Write $w(p)=B+C=1-2A(p)$ for the total stationary weight on the two asymmetric
outcomes $CD,DC$. By \eqref{SI:eq:AD} that weight is split evenly between them,
and likewise the remaining weight $1-w$ is split evenly between $CC$ and $DD$.
Substituting into \eqref{SI:eq:affine}, the expected payoff against a random
population collapses to a convex combination of the two diagonal averages,
\begin{equation}
  E(p,\Ran)\;=\;\bigl(1-w(p)\bigr)\,\frac{R+P}{2}
             \;+\;w(p)\,\frac{S+T}{2}.
  \label{SI:eq:master}
\end{equation}
A self-conjugate strategy can therefore differ from another only through the
single number $w$: it cannot bias the asymmetric play toward the $T$-role
rather than the $S$-role, only choose \emph{how often} the play is asymmetric
at all. Applying \eqref{SI:eq:master} to $\HOLD$ and $\SWAP$ and setting
\begin{equation}
  \xi(\eps)\;:=\;A(\SWAP)-A(\HOLD)
              \;=\;\tfrac12\bigl(w(\HOLD)-w(\SWAP)\bigr)\;>\;0,
  \label{SI:eq:xidef}
\end{equation}
every term cancels except the mixing weight, and we obtain the exact identity
\begin{equation}
  E(\HOLD,\Ran)-E(\SWAP,\Ran)\;=\;\xi(\eps)\,\bigl[\,(S+T)-(R+P)\,\bigr].
  \label{SI:eq:main}
\end{equation}
The constant $\xi$ is a property of the two strategies and of the law of $\Ran$
alone --- by \eqref{SI:eq:affine} it carries no dependence on the payoff matrix
whatsoever --- so the entire game-dependence of the difference sits in the
single factor $(S+T)-(R+P)$.

\paragraph{Values.}
Integrating $v_{CC}$ over $\Omega_\eps$ --- the $q_1$-integral in closed form,
since the numerator is independent of $q_1$ and the denominator is linear in
it, followed by high-order Gauss--Legendre quadrature in $(q_2,q_3,q_4)$ ---
gives
{\renewcommand{\arraystretch}{1.25}
\[
  \begin{array}{c|ccc}
     & \eps=10^{-3} & \eps=10^{-4} & \eps\to0\\\hline
    A(\SWAP) & 0.245943547164 & 0.245889648665 & 0.245883628111\\
    A(\HOLD) & 0.206347626011 & 0.205829111865 & 0.205770669828\\\hline
    \xi(\eps)& 0.039595921153 & 0.040060536800 & 0.040112958284
  \end{array}
\]}
with $\xi(0)=0.04011295828362\ldots$ (which is close to, but not equal to, $1/25$).
Thus $\SWAP$ reaches mutual cooperation more often against random opponents
($A(\SWAP)>A(\HOLD)$, and by the symmetry \eqref{SI:eq:AD} it also sits in mutual
defection just as much more often), whereas $\HOLD$ spends the larger share of
its time in the asymmetric states: $w(\HOLD)=0.5885>0.5082=w(\SWAP)$ as
$\eps\to0$. By \eqref{SI:eq:master} the second effect prevails precisely when the
asymmetric outcomes are on average worth more than the symmetric ones,
\[
  E(\HOLD,\Ran)\;>\;E(\SWAP,\Ran)
  \qquad\Longleftrightarrow\qquad
  S+T\;>\;R+P.
\]
Every snow-drift game satisfies $T>R$ and $S>P$, hence $S+T>R+P$, so a
population of random M1 strategies is biased towards $\HOLD$ from the outset.
Under the parameterization $R=1$, $P=0$, $S=u$, $T=1+v$ the identity \eqref{SI:eq:main} reads
$E(\HOLD,\Ran)-E(\SWAP,\Ran)=\xi(\eps)\,(u+v)$.\\

\renewcommand{\refname}{}
\section*{Supplementary References}
~\\[-2.5cm]
{{\setlength{\bibsep}{0\baselineskip}

}}
\newpage

\clearpage
\begin{center}
\includegraphics[width=\linewidth,height=0.68\textheight,keepaspectratio]{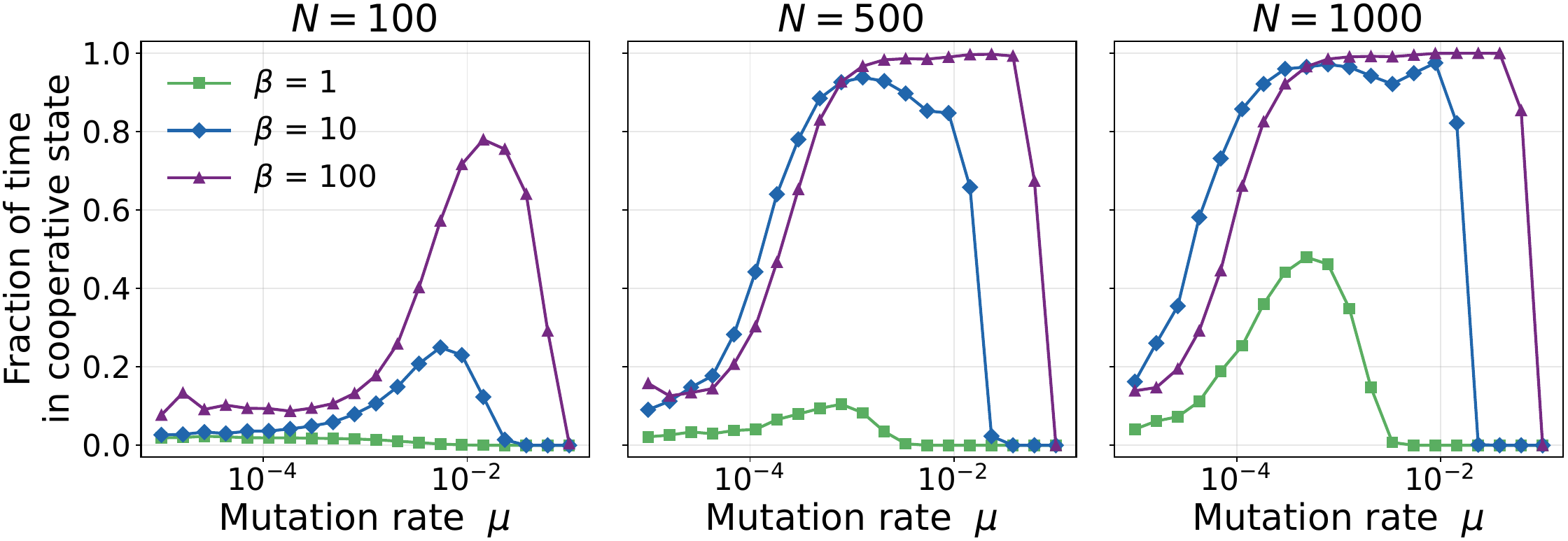}
\end{center}
\vspace{1.5ex}

\noindent {\small {\bf SI Figure 1: Exploring a costly donation game.} Individuals use M1 strategies and interact in
a donation game with a cost-to-benefit ratio of $0.95$ for three population sizes, $N=100, 500, 1000$ and three intensities of selection, $\beta=1,10,100$. The figure shows the time the population spends in the cooperative state as a function of the mutation rate, $\mu$. For larger population sizes and higher intensities of selection, there is a broad window of mutation rates that yield almost complete cooperation. At very high mutation rates, an error threshold leads to a sharp decline in cooperation.}

\clearpage
\begin{center}
\includegraphics[width=\linewidth,height=0.68\textheight,keepaspectratio]{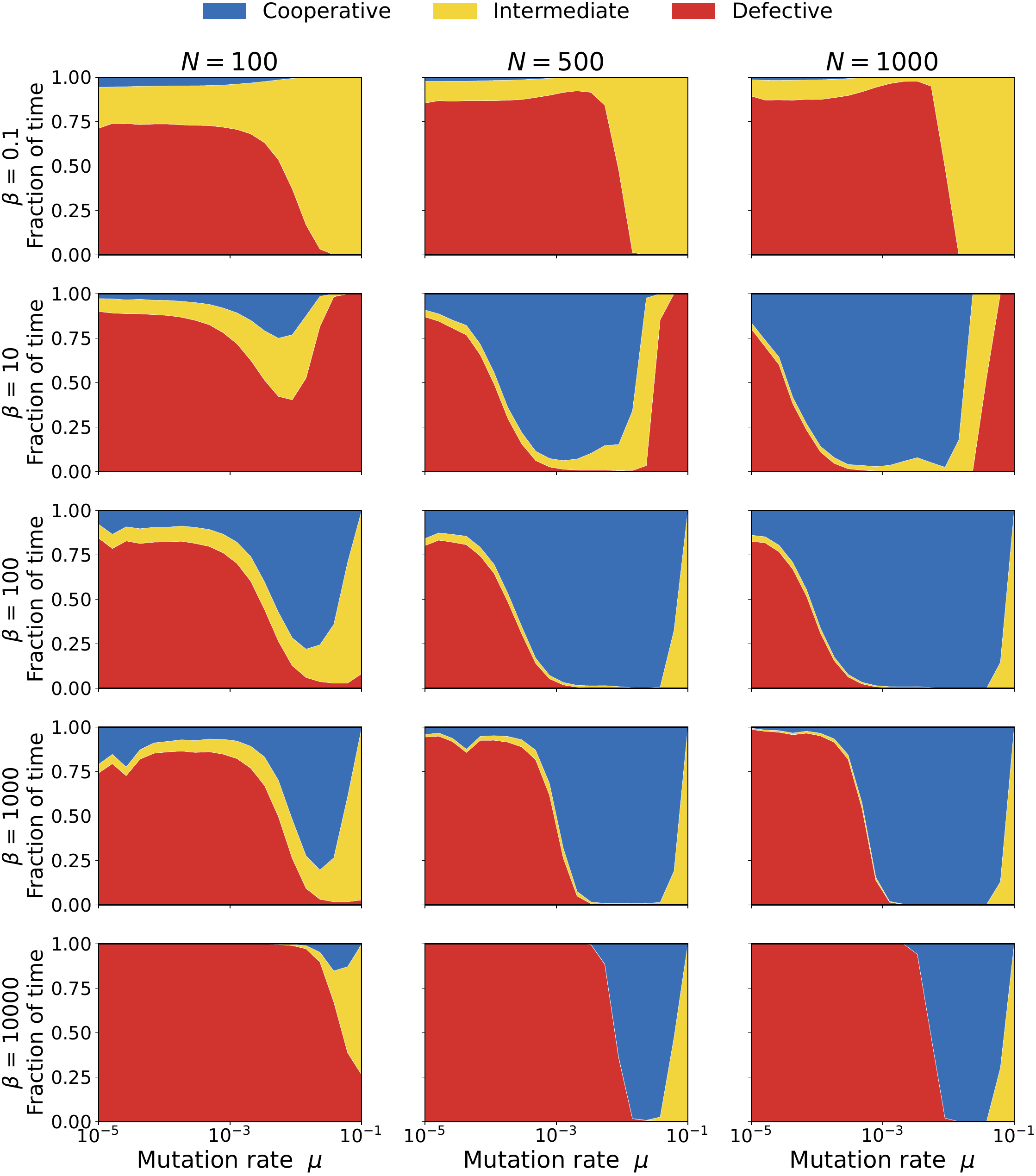}
\end{center}
\vspace{1.5ex}

\noindent {\small {\bf SI Figure 2: Occupancy of the three states.} The high dimensional, continuous state space is subdivided into three regions (as explained in the text): a cooperative region (blue), an intermediate region (yellow) and a defective region (red). The figure shows the fraction of time spent in each of those states as a function of mutation rate for three population sizes, $N$, and five intensities of selection, $\beta$. Note that very high selection intensities narrow the window of maximum cooperation. Standard parameters (see Text of SI).}

\clearpage
\begin{center}
\includegraphics[width=\linewidth,height=0.68\textheight,keepaspectratio]{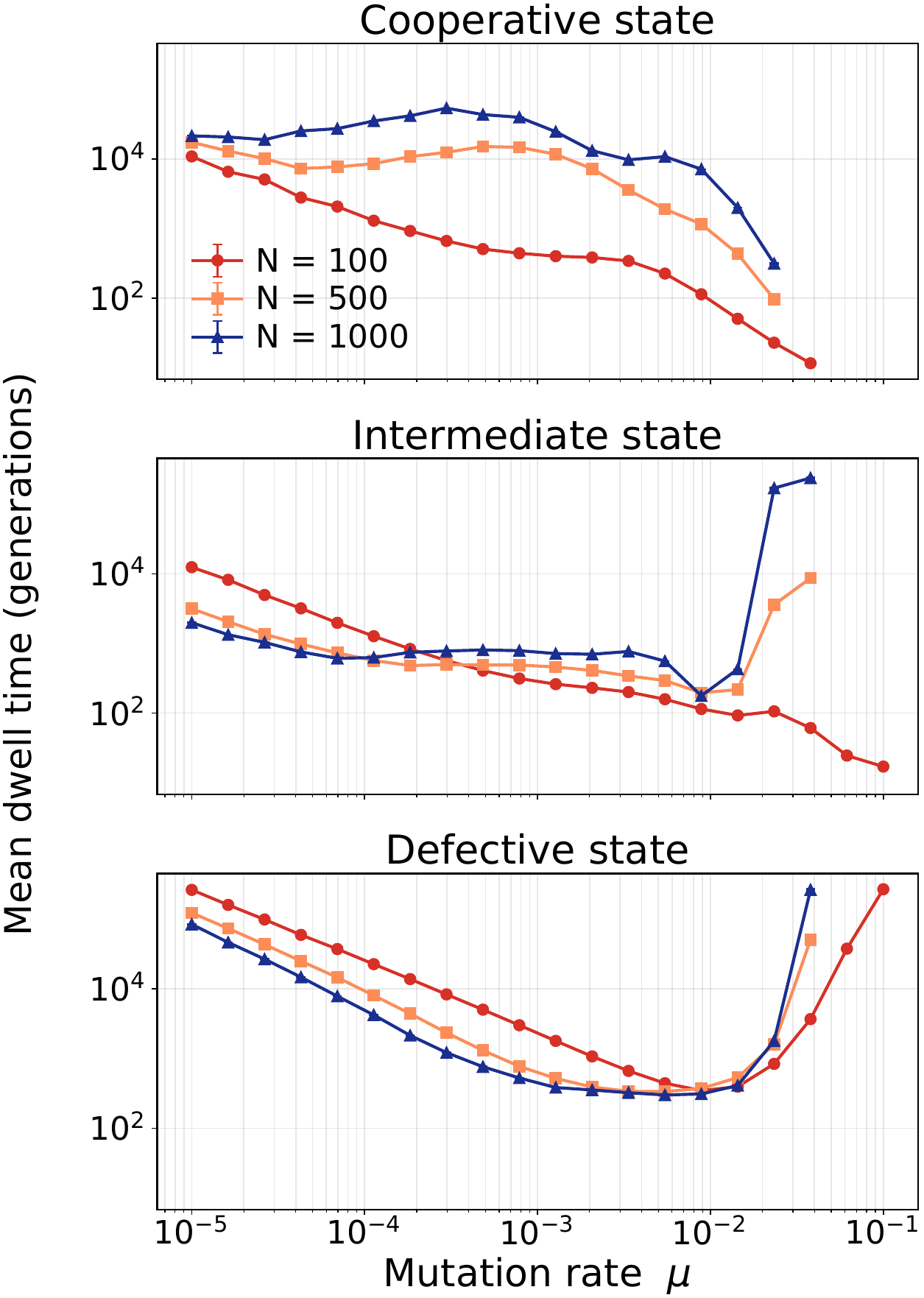}
\end{center}
\vspace{1.5ex}

\noindent {\small {\bf SI Figure 3: Dwell time in the cooperative, intermediate and defective state.}
For $N=100$, the mean dwell time in the cooperative state is a declining function of mutation rate. For $N=500$ and $N=1000$, the mean dwell time in the cooperative state is constant or increasing with mutation rate before declining at very high mutation rates. In contrast, for all population sizes, the mean dwell time in the defective state declines with mutation rate before crossing the error threshold when it rises sharply. Therefore, in large populations, increasing diversity (below the error threshold) favors cooperation.}

\clearpage
\begin{center}
\includegraphics[width=\linewidth,height=0.68\textheight,keepaspectratio]{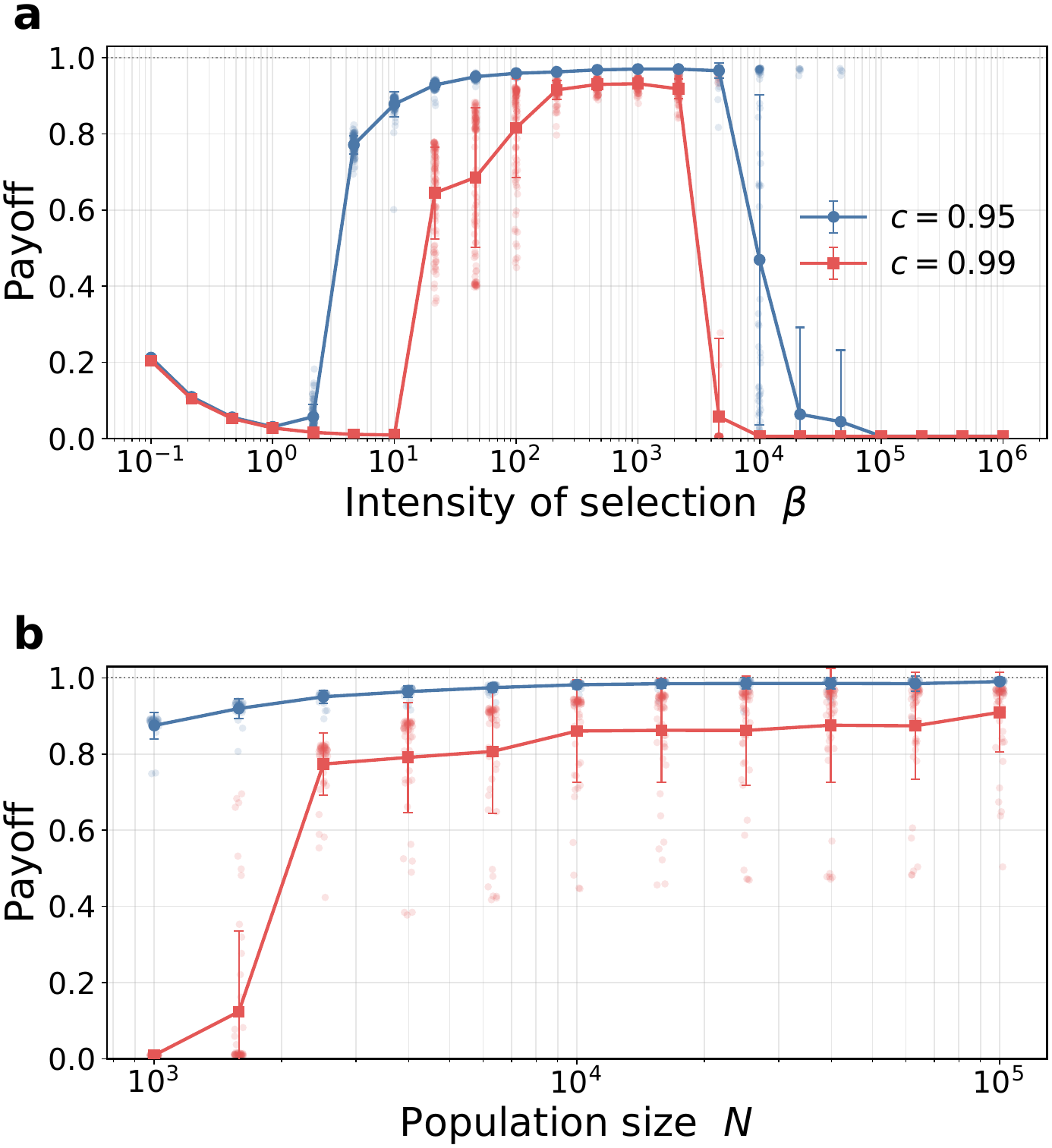}
\end{center}
\vspace{1.5ex}

\noindent {\small {\bf SI Figure 4: Efficiency versus intensity of selection and population size.}
Maximum efficiency requires an intermediate amount of selection but benefits monotonically from larger populations. The simulations use M1 strategies. {\bf a:} A broad range of intermediate selection intensities, $\beta$, yield maximum efficiency for resolving the donation game with very high cost values,  $c=0.95$ (blue) and $c=0.99$ (red). The selection intensity can be too weak and too strong. Parameters: $N=1000$, $\mu=10^{-2}$, 100 runs for each average.  {\bf b:} Efficiency increases with population size at least up to $N=10^5$, which is the maximum shown here. Parameters: $\beta=10$, $\mu=10/N$, 30 runs (c=0.95) and 50 runs  (c=0.99) for each average.}

\clearpage
\begin{center}
\includegraphics[width=\linewidth,height=0.68\textheight,keepaspectratio]{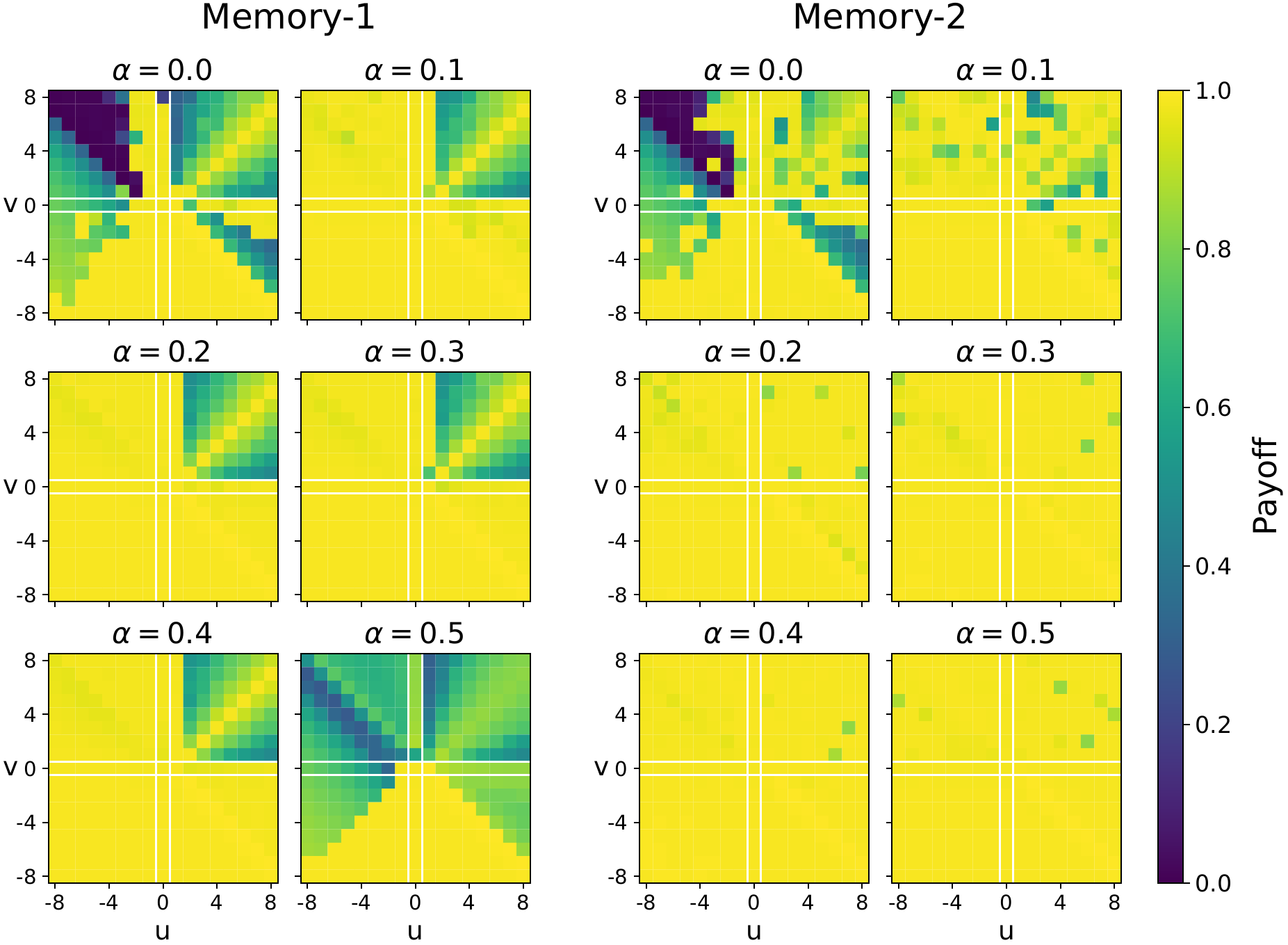}
\end{center}
\vspace{1.5ex}

\noindent {\small {\bf SI Figure 5. Varying the bias in the mutation process.}   Most of the relevant strategies exist near the boundary of the strategy space. The parameter, $\alpha \in [0,1/2]$, determines the bias of mutation toward the boundary. The random draw for each $p_i$ value has three outcomes: (i) with probability $\alpha$ we obtain $p_{min}$; (ii) with probability $1-2\alpha$ we obtain a uniformly distributed value in the interval $(p_{min},p_{max})$; (iii) with probability $\alpha$ we obtain $p_{max}$.  For $\alpha=0$ (no bias), many games remain unresolved. For $\alpha=0.1$, M1 strategies resolve many games (outside the Hold-trap), but M2 remains patchy. For $\alpha=0.2, 0.3, 0.4$, both M1 and M2 succeed. Using only binary strategies ($\alpha=0.5$) works well for M2 but not for M1.}

\clearpage
\begin{center}
\includegraphics[width=\linewidth,height=0.68\textheight,keepaspectratio]{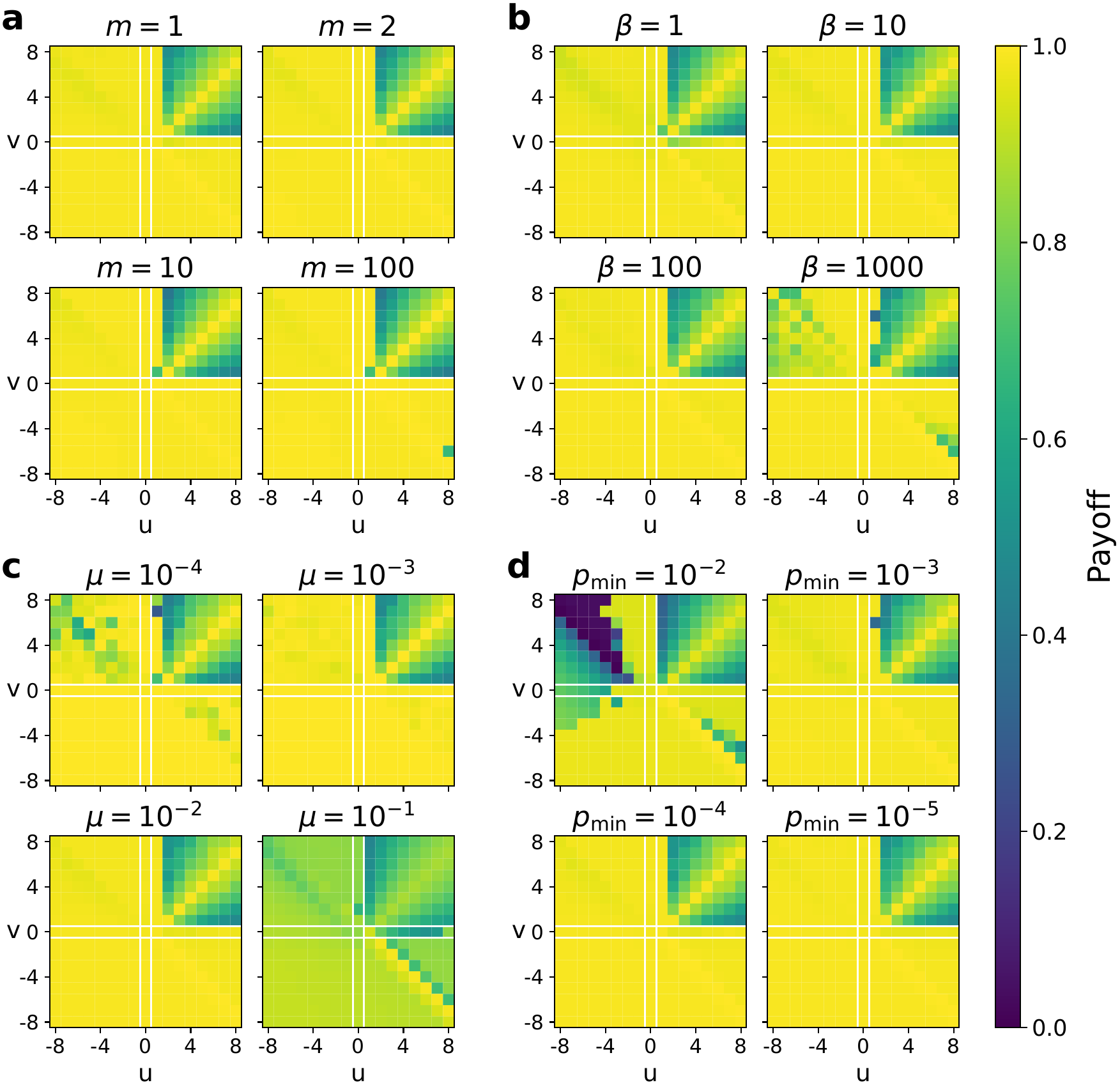}
\end{center}
\vspace{1.5ex}

\noindent {\small {\bf SI Figure 6. Varying key parameters.} We study the effect of variation in key parameters using M1 strategies. {\bf a:} Varying the number of interaction partners, $m=1, 2, 10, 100$, has essentially no effect. Being able to simulate the process for $m=1$ is a dramatic speed up in computation time. {\bf b:} We observe high efficiency for a broad range of $\beta$ values. For $\beta=1000$ we lose efficiency in some games. {\bf c:} High efficiency is obtained for intermediate mutation rates, $\mu$. {\bf d:} When the minimum noise level, $\epsilon$, gets too large ($\epsilon \approx 0.01$), we lose efficiency in some games, especially in the Prisoner's Dilemma.}

\clearpage
\begin{center}
\includegraphics[width=\linewidth,height=0.68\textheight,keepaspectratio]{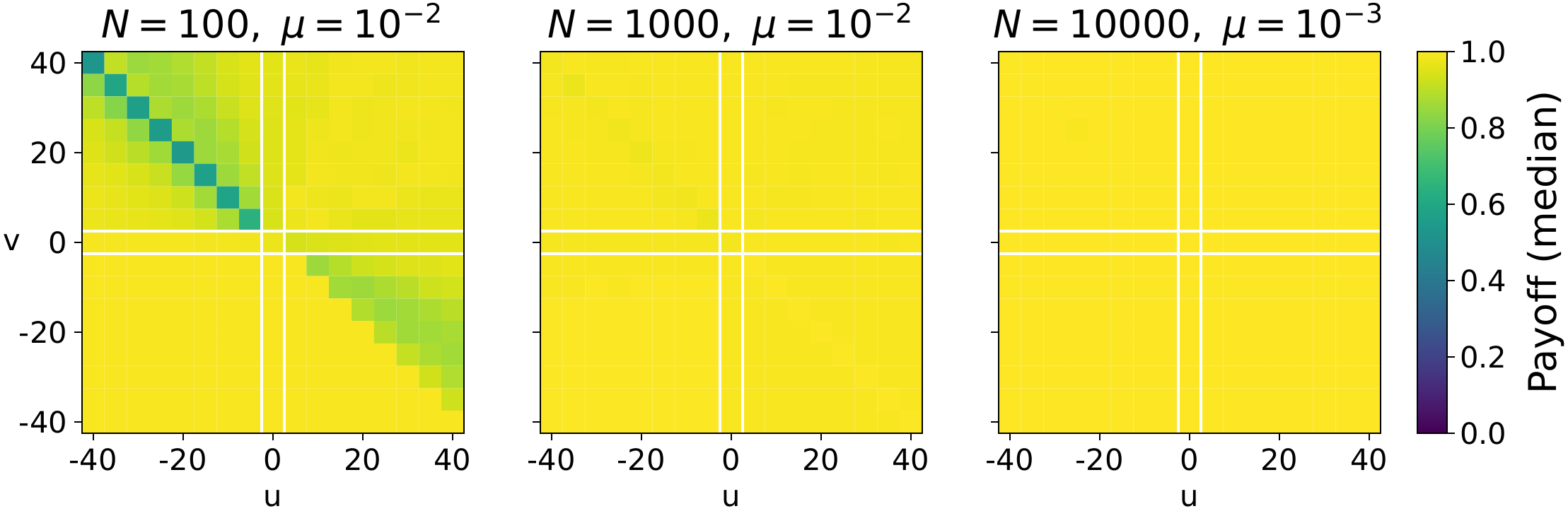}
\end{center}
\vspace{1.5ex}

\noindent {\small {\bf SI Figure 7. Zooming out.}  The payoff parameters $u$ and $v$ vary from $-40$ to $+40$ in steps of 5. M2 strategies resolve games with such extreme payoff values. Population sizes are $N=100, 1000, 10000$ as shown. Each point in the heat map is the median over 10 independent runs. Parameters: mutation rate $\mu=10^{-2}$ for $N=100$ and $N=1000$, and $\mu=10^{-3}$ for $N=10^4$, as indicated above each panel; time $t=10^7$; otherwise standard parameters.}

\clearpage
\begin{center}
\includegraphics[width=\linewidth,height=0.68\textheight,keepaspectratio]{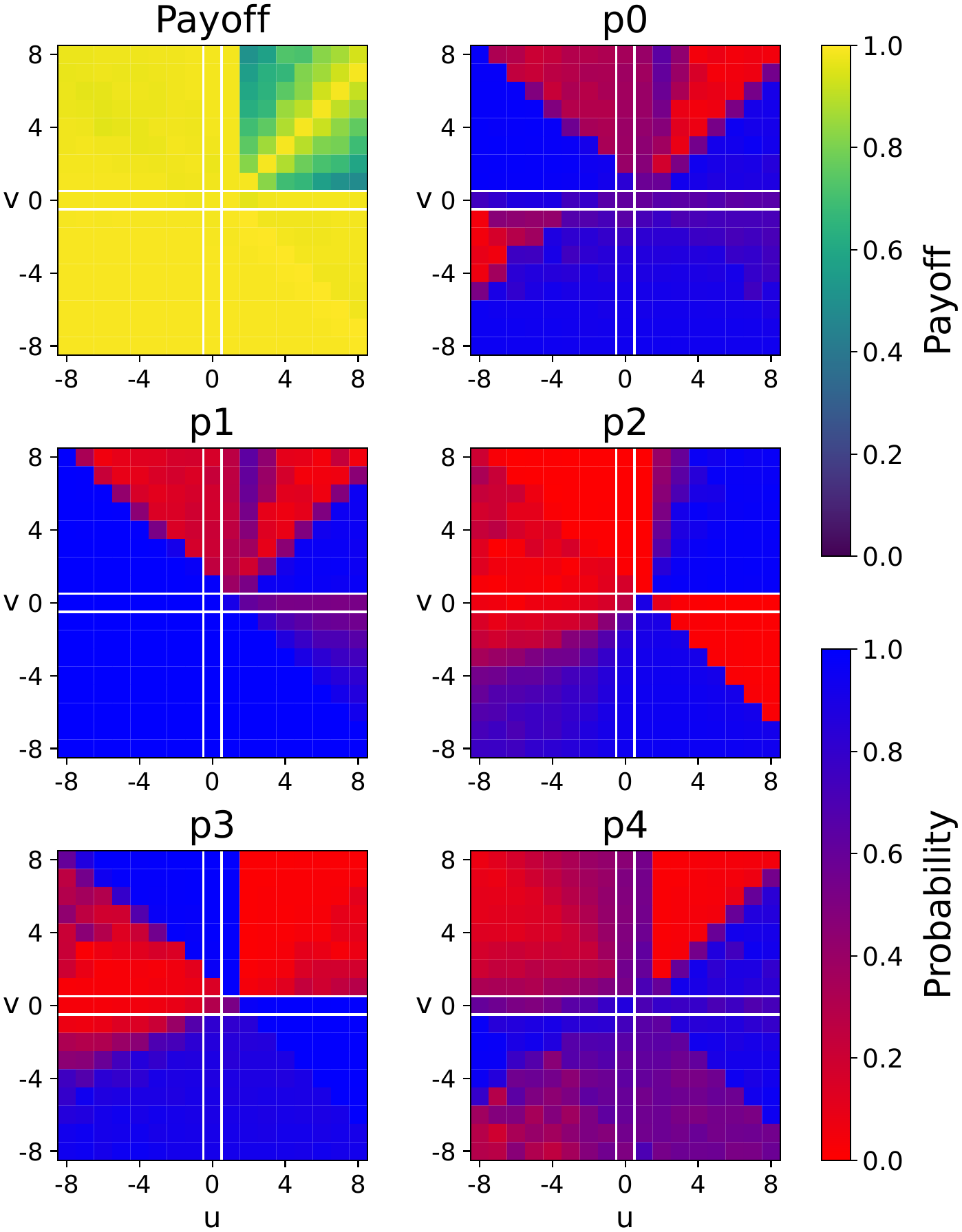}
\end{center}
\vspace{1.5ex}

\noindent {\small {\bf SI Figure 8. Finitely repeated games.}  For finitely repeated games, M1 strategies are given by 5 parameters: $(p_0,p_1,p_2,p_3,p_4)$. We consider a probability of 0.001 that the game will end after each round. We obtain very similar outcomes as for the infinitely repeated game. Standard parameters.}

\clearpage
\begin{center}
\includegraphics[width=\linewidth,height=0.68\textheight,keepaspectratio]{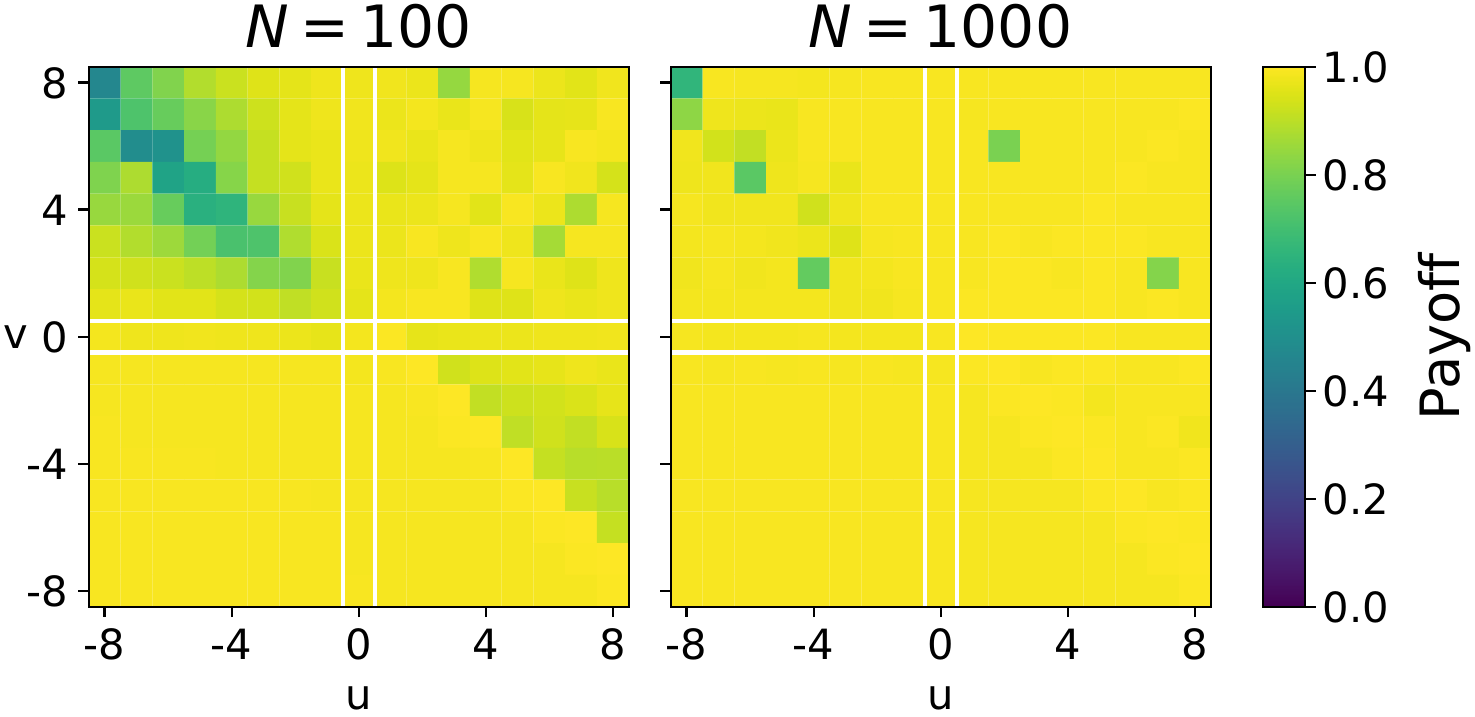}
\end{center}
\vspace{1.5ex}

\noindent {\small {\bf SI Figure 9. Asynchronous updating.} We study M2 strategies in a process with asynchronous updating (see text of SI). The figure shows efficiency as a function of the payoff
 parameters, $u$ and $v$, for two population sizes, $N=100$ and $N=1000$. 
 Color gives the mean payoff, from $0$ (minimum efficiency, dark blue) to $1$
 (maximal efficiency, yellow); white lines mark $u=0$ and $v=0$. All other parameters
 are fixed ($m=2$, mutation rate $\mu=10^{-2}$, $\alpha=1/3$, $p_{\min}=10^{-4}$,
 $\beta=10$). The larger population is more efficient overall (mean $0.98$ vs. $0.95$);
 the two differ most in the Prisoner's Dilemma quadrant ($u<0$, $v>0$), where random drift in
 the small population destabilizes cooperation.}

\end{document}